\documentclass[usenatbib,a4paper,times,fleqn]{mnras}
\usepackage{graphicx,natbib,times,amssymb,amsmath,pstricks}
\usepackage{aas_macros}
\usepackage{bm,url}
\usepackage{textcomp}
\usepackage{breakurl}
\defcitealias{lacouretal16}{L16}

\title[Circumbinary dynamics in the HD142527 disc]{Circumbinary, not transitional: On the spiral arms, cavity, shadows, fast radial flows, streamers and horseshoe in the HD142527 disc}

\author[Price et al.]{\parbox{\textwidth}{Daniel J. Price$^{1}$\thanks{daniel.price@monash.edu}, Nicol\'as Cuello$^{2,3,4}$, Christophe Pinte$^{1,5}$, Daniel Mentiplay$^{1}$, Simon Casassus$^{6,3}$, Valentin Christiaens$^{6,3}$, Grant M. Kennedy$^{7}$, Jorge Cuadra$^{2,3,4}$, Sebastian Perez M.$^{6,3}$, Sebastian Marino$^{7}$, Philip J. Armitage$^{8,9}$, Alice Zurlo$^{6,3}$, Attila Juhasz$^7$, Enrico Ragusa$^{10}$, Guillaume Laibe$^{11}$ and Giuseppe Lodato$^{10}$} \vspace{0.2cm} \\
$^{1}$Monash Centre for Astrophysics (MoCA) and School of Physics and Astronomy, Monash University, Clayton Vic 3800, Australia\\
$^{2}$Instituto de Astrof\'isica, Pontificia Universidad Cat\'olica de Chile, Santiago, Chile\\
$^{3}$Millennium Nucleus `Protoplanetary discs', Chile \\
$^{4}$N\'ucleo Milenio de Formaci\'on Planetaria (NPF), Chile \\
$^{5}$Univ. Grenoble Alpes, CNRS, IPAG / UMR 5274, F-38000 Grenoble \\
$^{6}$Departamento de Astronom\'ia, Universidad de Chile, Casilla 36-D, Santiago, Chile\\
$^{7}$Institute of Astronomy, University of Cambridge, Madingley Rd, Cambridge, CB3 0HA, UK \\
$^{8}$JILA, University of Colorado \& NIST, UCB 440, Boulder, CO 80309-0440, USA \\
$^{9}$Department of Astrophysical and Planetary Sciences, University of Colorado, 391 UCB, Boulder, CO 80309-0391, USA \\
$^{10}$Dipartimento di Fisica, Universit\`a Degli Studi di Milano, Via Celoria, 16, Milano, I-20133, Italy \\
$^{11}$Univ Lyon, Univ Lyon1, Ens de Lyon, CNRS, Centre de Recherche Astrophysique de Lyon UMR5574, F-69230, Saint-Genis-Laval, France
}
\pagerange{\pageref{firstpage}--\pageref{lastpage}} \pubyear{2017}

\date{}
\begin{document}
\label{firstpage}
\bibliographystyle{mnras}
\maketitle

\begin{abstract}
 We present 3D hydrodynamical models of the HD142527 protoplanetary disc, a bright and well studied disc that shows spirals and shadows in scattered light around a 100~au gas cavity, a large horseshoe dust structure in mm continuum emission, together with mysterious fast radial flows and streamers seen in gas kinematics. By considering several possible orbits consistent with the observed arc, we show that all of the main observational features can be explained by one mechanism --- the interaction between the disc and the observed binary companion. We find that the spirals, shadows and horseshoe are only produced in the correct position angles by a companion on an inclined and eccentric orbit approaching periastron --- the `red' family from Lacour et al. (2016). Dust-gas simulations show radial and azimuthal concentration of dust around the cavity, consistent with the observed horseshoe. The success of this model in the HD142527 disc suggests other mm-bright transition discs showing cavities, spirals and dust asymmetries may also be explained by the interaction with central companions.
\end{abstract}


\section{Introduction}
 Around the young star HD142527 lies an enigmatic and spectacular protoplanetary disc. Cycle 0 observations with the Atacama Large Millimetre/submillimetre Array (ALMA) by \citet{casassusetal13} revealed a `horseshoe' of dust continuum emission (first detected by \citealt{ohashi08}) surrounding a $\sim$ 90--140~au central cavity (cavities being the defining feature of so-called `transitional' or `transition discs'; \citealt{strometal89,espaillatetal14}). The first ALMA observations also revealed diffuse CO gas inside the cavity alongside mysterious `streamers' or `filaments' seen in HCO+ emission, thought to indicate flow across a planet-induced gap \citep{casassusetal13}. 
 
  The star itself is a Herbig Fe star of spectral type F6 IIIe (M$\approx$1.8~M$_{\odot}$) at a distance of $156^{+7}_{-6}$ pc \citep{gaiaetal16} in the Sco-Cen association (\citealt{billeretal12,mendigutiaetal14}). Modelling of the Spectral Energy Distribution (SED) suggested a disc gap between 30 and 130 au \citep{verhoeffetal11}, confirmed by the initial ALMA observations \citep{casassusetal13}. Earlier mid-infrared observations \citep{van-boekeletal04,fujiwaraetal06} and more recent scattered light images \citep{avenhausetal17} found a small inner disc inside the cavity of $\sim$10 au in radius. The high accretion rate onto the central star ($\approx 2 \times 10^{-7}$ M$_{\odot}$/yr; \citealt{garcia-lopezetal06,mendigutiaetal14}) implies that the small inner disc must be refilled from the outer disc, most likely in an episodic manner \citep{casassusetal13}.

 Even before ALMA, the disc around HD142527 had proved spectacular, with a spiral arm detected at $R \gtrsim 100$~au in scattered light observations in the near IR by \citet{fukagawaetal06}. Further observations revealed a wealth of spiral structure, including small near IR spirals at the edge of the cavity \citep{casassusetal12,canovasetal13,avenhausetal14,avenhausetal17}, a counterpart to the \citet{fukagawaetal06} spiral seen in the CO emission in further ALMA observations, as well as two further large scale ($\sim$ 500 au) spiral arms by \citet{christiaensetal14}.

  \citet{fukagawaetal06} first suggested that the spirals might be caused by an inner companion, with the presence of a 0.1--0.4 M$_{\odot}$ companion with a projected separation of $\sim 13$~au confirmed by \citet{billeretal12} using Sparse Aperture Masking (SAM) with the NACO instrument on the Very Large Telescope (VLT). More recent observations have refined both the orbit and the companion mass \citep{closeetal14,lacouretal16} (hereafter \citetalias{lacouretal16}).
  
  Scattered light images also revealed shadows cast across the outer disc \citep{fukagawaetal06,avenhausetal14}. \citet{mpc15} satisfyingly reproduced the observed illumination pattern by assuming a compact $\sim~10$ au inclined and thin inner disc casting shadows on the outer disc, with a relative inclination of $\sim 70^{\circ}$. This is consistent with the estimated inclination of the inner binary with respect to the disc \citepalias{lacouretal16}.
 
 Perhaps the greatest surprise were the complexities found in detailed kinematic studies using CO(6--5) molecular line data by \citet{casassusetal15}, with evidence for a warped disc and near free-fall motions within the central cavity, suggested to be related to theoretical models of `disc tearing' in warped accretion discs \citep{nkp13,npn15}.
 
Understanding the origin of features in the HD142527 disc is important because several of these appear common to a broad class of discs. In particular, dust horseshoes or rings surrounding the central cavities in mm-bright transition discs appear to be widespread (\citealt{van-der-mareletal16,canovasetal16}; see review by \citealt{casassus16}), with horseshoes normally interpreted as `dust traps' caused by a vortex at the edge of a planet-induced gap \citep[e.g.][]{pinillaetal12,van-der-mareletal13,baruteauzhu16,marinoetal15}. Spirals are also observed in an increasing number of discs \citep{garufietal13,benistyetal15,perezetal16,benistyetal17}. Thus, unlocking the mystery of HD142527 may help to shed light on the origin of these features in this wider class of discs.

Our approach is to perform 3D hydrodynamical and dust-gas simulations of the disc-binary interaction, to try to explain the observed features in the HD142527 disc. We describe the key observations in Section~\ref{sec:observations}, methods and initial conditions in Section~\ref{sec:simulations}. Results are in Section~\ref{sec:results}. We discuss in Section~\ref{sec:discussion} and summarise in Section~\ref{sec:summary}.

\begin{figure*}
\begin{center}
\includegraphics[width=0.96\columnwidth]{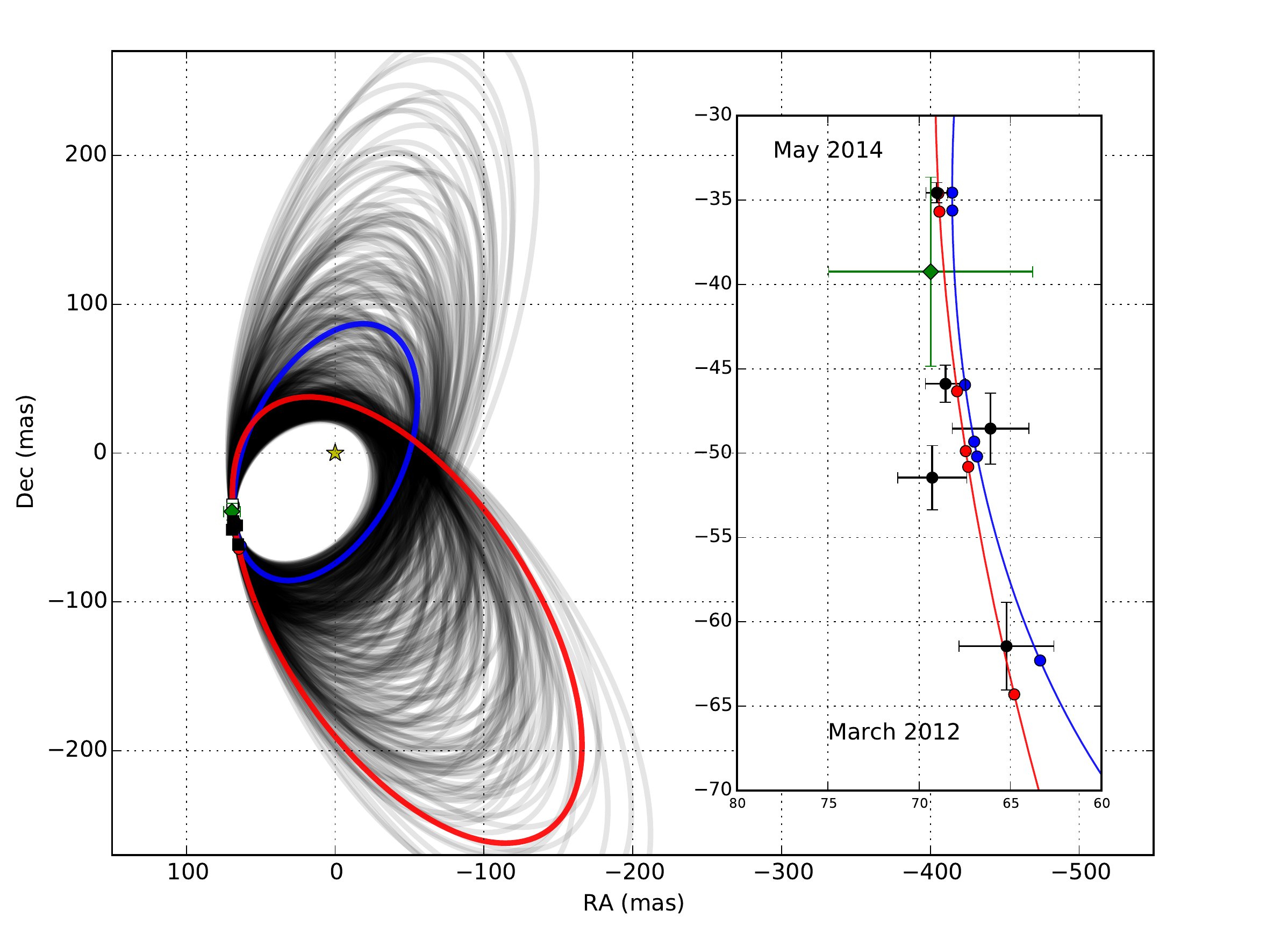}
\includegraphics[width=\columnwidth]{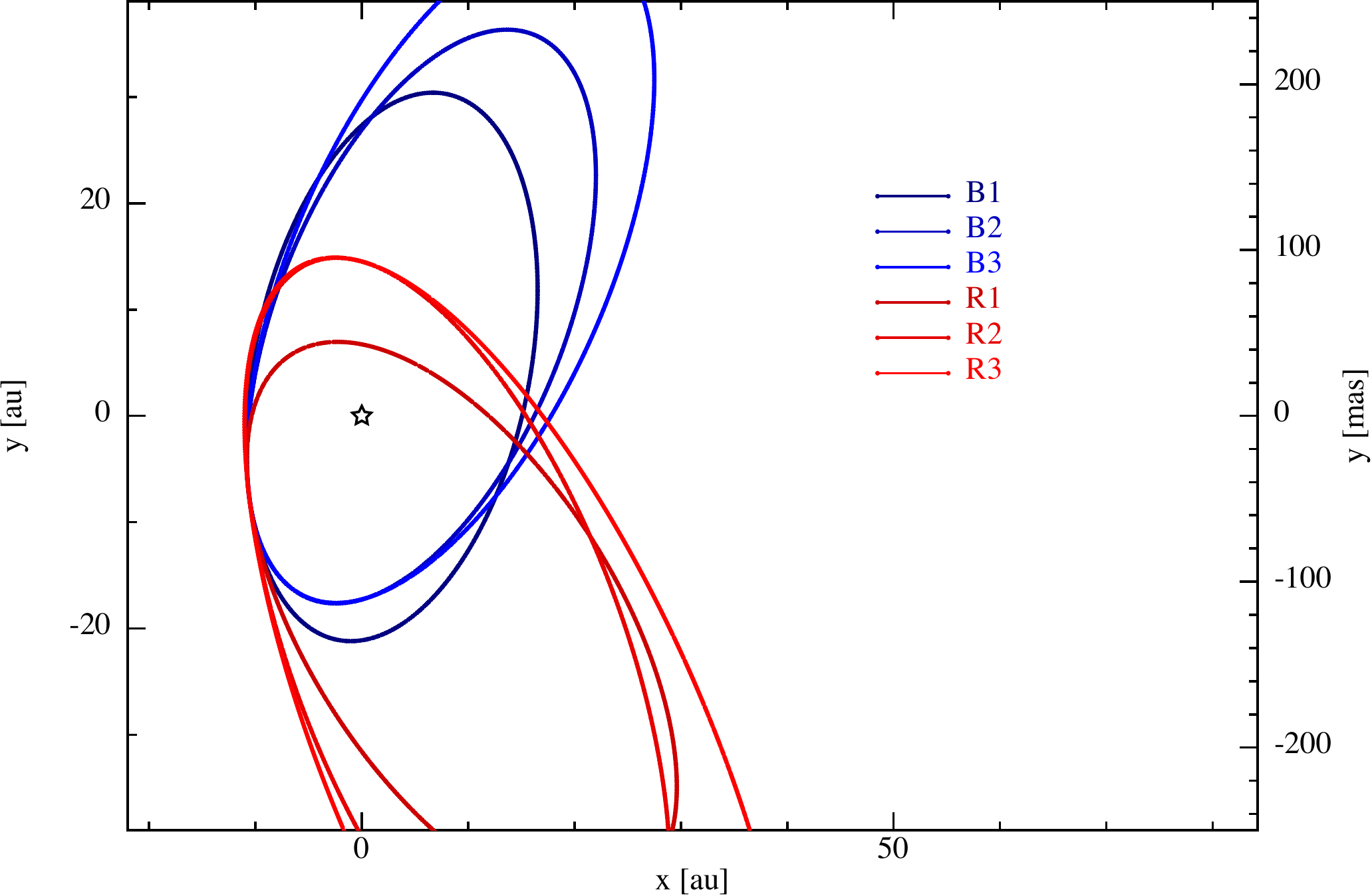}
\caption{Binary orbit. Left: Orbital fits for HD142527B (Credit: \citealt{lacouretal16}, reproduced with permission $\copyright$ESO). Right: Selected trial orbits for HD142527B used in this paper, corresponding to the orbital elements listed in Table~\ref{tab:orbits}. We assume the GAIA distance of 156pc. The star gives the location of the primary. Motion is clockwise.} 
\label{fig:orbits}
\end{center}
\end{figure*}

\section{Observational constraints}
\label{sec:observations}
 The sheer volume of data collected on HD142527 provides tight constraints on theoretical models, and it seems optimistic to assume that a single model could explain multiple observed features simultaneously. From a dynamical perspective the main puzzles are:
\begin{enumerate}
\item Shadows. The close agreement found by \citet{mpc15} between their radiative transfer model and the scattered light images means these shadows are almost certainly cast by an inner circumprimary disc inclined to the outer disc plane by $\sim 70^{\circ}$. The size of such a disc would be consistent with the infrared observations \citep{verhoeffetal11}. The constraints from ALMA CO observations of the inner disc suggests that it is small and with unusual (non-Keplerian) kinematics between the inner and outer disc \citep{perezetal15,casassusetal15}.
\item Fast radial flows. \citet{casassusetal15} suggested an explanation for fast radial flows in terms of disc tearing by an inclined inner binary \citep[e.g.][]{nkp13}. However, subsequent simulations by Dunhill and collaborators (reported in \citealt{cuadra16}) found that such a binary tended to simply break the disc into two distinct sections, as found by \citet{flp13} and consistent with expectations of warp dynamics in thick ($H/R \sim 0.1$) protoplanetary discs. \citet{rca14} also found that the fast radial flows were better explained by free-fall radial velocities rather than a warp, suggesting material is somehow able to shed angular momentum and plunge into the central regions. Those authors proposed gravitational torques from giant planets or brown dwarfs as a possible solution.
\item Spiral arms. Spirals are seen immediately outside the cavity in both scattered light \citep{fukagawaetal06,rodigasetal14,avenhausetal14,avenhausetal17} and in CO emission \citep{christiaensetal14}. These are usually attributed to the presence of either companions orbiting interior or exterior to the arms \citep{dongetal15}, to a gravitationally unstable disc \citep{dipierroetal15a}, or to some combination of both \citep{pohletal15}. \citet{quillen06} offered an explanation of similar spirals seen in the disc around HD100546 in terms of a precessing, warped disc driven by misaligned embedded protoplanets. \citet{montesinosetal16} also showed that spiral structure could be induced by temperature fluctuations caused by shadows.
\item Cavity. The origin of central cavities in transitional discs is not yet certain \citep{williamscieza11}. Traditionally central holes were thought to arise from either photoevaporation of gas by the central star or depletion of gas and/or dust due to formation of planetary mass companions \citep{williamscieza11,andrewsetal11,espaillatetal14,owen16}. \citet{zhuetal12} found that dust filtering by giant planets \citep[e.g.][]{riceetal06} could partially explain the mm-dust holes seen in many transition discs, but that the depletion in small particles was insufficient to explain the near-IR deficit in the Spectral Energy Distribution (SED). An alternative possibility is the tidal truncation of the disc from a central binary \citep{artymowiczlubow94}. While a low mass companion has been found in HD142527 (see below), previous authors (e.g. \citetalias{lacouretal16}) have assumed that the projected orbital separation of $\sim 13$ au is too small to tidally truncate the disc out to 100 au --- the size of the observed CO cavity \citep{perezetal15,mutoetal15,boehleretal17}.
\item Dust horseshoe. Currently the main accepted model for producing dust horseshoes in mm-bright transition discs involves dust trapping by a gap-edge vortex \citep{pinillaetal12,lyralin13}. \citet{ataieeetal13} considered an alternative model where dust horseshoes could be produced by eccentric circumbinary cavities, but dismissed this model based on their simulations. However, dust evolution was added only in post-processing which neglects the role of tidal torques on the dust and backreaction, leading to potentially misleading conclusions. Recently, \citet{ragusaetal17} showed that central binaries can produce both rings and horseshoes in dust emission. Indeed, eccentricities around the cavity edge are a common feature in hydrodynamical simulations of circumbinary discs \citep{kleydirksen06,farrisetal14,rlp16}. \citet{ragusaetal17} showed that more massive companions produce progressively more asymmetric structures, with simulated ALMA observations closely matching observed discs.
\item Gap-crossing filaments. \citet{casassusetal13} attributed these to flows of gas through a planetary gap, as observed commonly in simulations of gap-opening by embedded planets in discs (e.g. \citealt{brydenetal99,lubowdangelo06}). However, it is not clear how the `filaments' might be obviously related to either the companion or any other putative planets, nor how they relate to the fast radial flows, or whether they may be illumination effects due to shadowing from the tilted inner disc.
\end{enumerate}

 A clue to solving the puzzle is that all of the above features may in principle be caused by the interaction with an (inclined) central binary \citep{fukagawaetal06}. The main issue is how a binary with a $\sim 13$ au projected separation could carve a $\sim 90$--$140$ au cavity. This led \citetalias{lacouretal16} to conclude that the cavity could not be created by the binary. However, the orbit is poorly constrained, with best fitting orbits from \citetalias{lacouretal16} suggesting significant eccentricity.
 

\section{Numerical methods}
\label{sec:simulations}
 We perform 3D hydrodynamics simulations of the disc-binary interaction using the {\sc Phantom} smoothed particle hydrodynamics (SPH) code (\citealt{priceetal18}; for reviews of SPH see \citealt{monaghan05,price12}). We also perform several models with dust and gas to see if we can simultaneously produce the dust horseshoe. Our approach to dust-gas modelling is similar to that used in our previous papers \citep{dipierroetal15,dipierroetal16}, where we model the dust using a separate set of particles coupled to the gas via a drag term \citep{laibeprice12,laibeprice12a,priceetal18}.

\subsection{Initial conditions}

\subsubsection{Binary}
 
 
 We model the binary using sink particles that interact with the gas disc only via gravity and by accreting gas \citep*{bbp95}. The sink particles are free to migrate and also change mass and orbital parameters due to interaction with the gas disc. We set the primary mass to 1.8 M$_{\odot}$ (from the spectral type), and constrained by the total mass estimate of $\sim 2.1 \pm 0.2$ M$_{\odot}$ from the Keplerian motion of the outer disc \citep{casassusetal15} we set the companion mass to 0.4 M$_{\odot}$ (the latest observational estimates based on spectral fitting are in the range 0.2--0.4 M$_{\odot}$; \citealt{christiaensetal18}). Since our chosen mass of 0.4~M$_\odot$ is at the higher end of this range, we also performed an additional set of calculations employing lower mass companions. We found that our results are not strongly sensitive to the secondary mass, with companion masses as low as 0.1M$_\odot$ producing the same cavity but with lower amplitude perturbations around the cavity edge. The change in mass due to accretion is negligible in our simulations.
 
 We fix the accretion radii for both sinks to $1$~au in order to resolve the circumprimary disc, if present.
 
 \begin{table}
\begin{tabular}{l|c|c|c|c|c|c|clcl}
\hline
Orbit & $a$ & $e$ & $i$ & $\Omega$ & $\omega$ & $f$ & $P$ & $\theta$ \\
\hline
B1 & 26.5 & 0.24 & 119.9 & 349.7 & 218.0 & 25.93 &  91.8  & 40.4 \\ 
B2 & 28.8 & 0.40 & 120.4 & 340.3 & 201.5 & 33.78 & 104 & 39.6 \\ 
B3 & 34.3 & 0.50 & 119.3 & 159.2 & 19.98 & 35.04 &  135 & 80.7 \\ 
R1 & 31.4 & 0.74 & 131.3 & 44.95 & 27.88 & 249.3 &  118 & 43.1 \\ 
R2 & 38.9 & 0.61 & 120.3 & 19.25 & 354.0 & 268.3 & 164 & 45.3 \\ 
R3 & 51.3 & 0.70 & 119.3 & 201.4 & 173.3 & 270.4 & 247 & 76.3 \\ 
\hline
\end{tabular}
\caption{Orbital elements for HD142527B for 6 trial orbits, drawn from fits to the observed arc with {\sc imorbel}. From left to right: semi-major axis ($a$,~au), eccentricity ($e$), inclination ($i$, deg), position angle of ascending node ($\Omega$, deg; East of North), argument of pericentre ($\omega$, deg), true anomaly ($f$, deg), orbital period ($P$, yrs) and relative angle between disc and binary ($\theta$, deg).}
\label{tab:orbits}
\end{table}

\begin{figure*}
\begin{center}
\includegraphics[width=\textwidth]{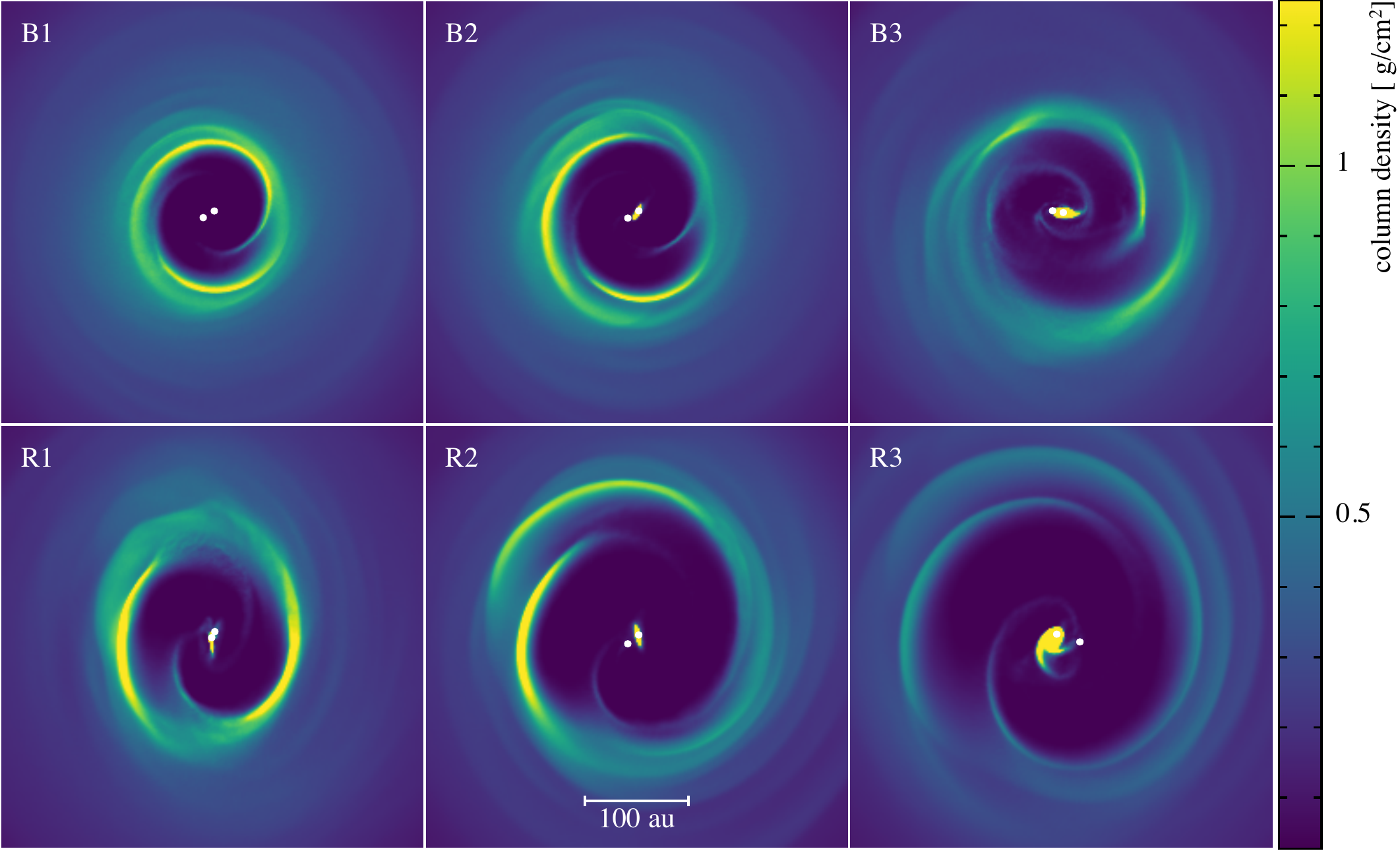}
\caption{Gas surface densities after 20 orbital periods of the binary for the calculations with $R_{\rm in} = 50$ au, showing the initial dynamical carving of the cavity for binary orbits shown in Figure~\ref{fig:orbits} and listed in Table~\ref{tab:orbits}. Top row shows the `blue' orbits (binary just past periastron) while bottom row corresponds to the `red' orbits (binary approaching periastron) from \citetalias{lacouretal16}. Cavity size scales with apastron separation, with more eccentric binaries (R2, R3) producing cavity sizes consistent with those observed in HD142527. Transient circumprimary discs are visible in all calculations except B1.} 
\label{fig:cavity}
\end{center}
\end{figure*}

\subsubsection{Binary orbit}
 After a few trial simulations with binaries of various semi-major axes and eccentricities, and from previous attempts at modelling HD142527 \citep[see][]{cuadra16}, we realised the importance of the known observational constraints on the orbit. To this end we employed {\sc imorbel}\footnote{\url{http://github.com/drgmk/imorbel}} to fit the orbit, written by Tim Pearce and Grant Kennedy \citep{pwk15}. The same code was also used in \citetalias{lacouretal16}. 
 
 Rather than taking the orbital fits directly from \citetalias{lacouretal16}, we produced a revised set of orbital fits using the GAIA distance measurement and with our assumed primary and secondary masses. {\sc imorbel} produces a plot similar to the one shown in figure 5 of \citetalias{lacouretal16}, from which one may interactively select orbits with given parameters that match the observational constraints. Our guiding wisdom in selecting trial orbits was i) to examine orbits similar to the `blue' and `red' orbit families found in the Monte Carlo fitting shown in Figure 6 of \citetalias{lacouretal16}; and ii) to ensure an apoastron separation for the binary large enough to plausibly explain the cavity size.

 Table~\ref{tab:orbits} lists the orbital elements used for the six representative orbits shown in this paper, with the resultant orbits shown in Figure~\ref{fig:orbits}. For clarity we show three representative ``blue" orbits (B1, B2, B3) and three ``red" orbits (R1, R2, R3) listed in order of increasing semi-major axis. The most tightly constrained parameter is the inclination, which is $i = 120^\circ$ for all but one of the orbits. We choose eccentricities ranging from 0.24 to 0.7. The two classes of orbits are `families' only in a projected sense, as the values of $\Omega$ and $\omega$ (and hence the orientation perpendicular to the sky plane) change dramatically. The main similarity between each family is simply the location of the periastron in position angle on the sky and thus whether the binary is approaching (red) or receding (blue) from periastron. The last column in Table~\ref{tab:orbits} gives the angle between the disc and binary angular momentum vectors.
 
 The Monte Carlo fitting performed by \citetalias{lacouretal16} found semi-major axes $11 < a < 40$ au, eccentricities $e=0.5 \pm 0.2$ and inclinations $125 \pm 15$ degrees to within 1$\sigma$ probability. Thus all our chosen orbits except R3 fall are probable to within 1$\sigma$ of these constraints.

 Finally, we adopted slightly different conventions for the orbital elements to those used in \citet{pwk15} and \citetalias{lacouretal16}. In particular, we removed the assumption by \citet{pwk15} that the angle between the binary and the sky is restricted to be less than $180^\circ$ (thus flipping the observer from $+z$ to $-z$ depending on the orbit). Since the absolute orientation of the disc and binary are irrelevant in SPH, for convenience we defined the observer to be viewing the disc down the $z-$axis (i.e. from $z=\infty$). That is, we tilted both the disc and binary in our initial setup. This simplified our analysis since replicating the observers view of HD142527 simply meant making an $x$-$y$ plot in our computational coordinates. Finally, we adapted the setup routine in {\sc Phantom} to use orbital elements from {\sc imorbel} in the form given in Table~\ref{tab:orbits}, such that the public codes are mutually compatible.
 
\begin{figure*}
\begin{center}
\includegraphics[width=0.9\textwidth]{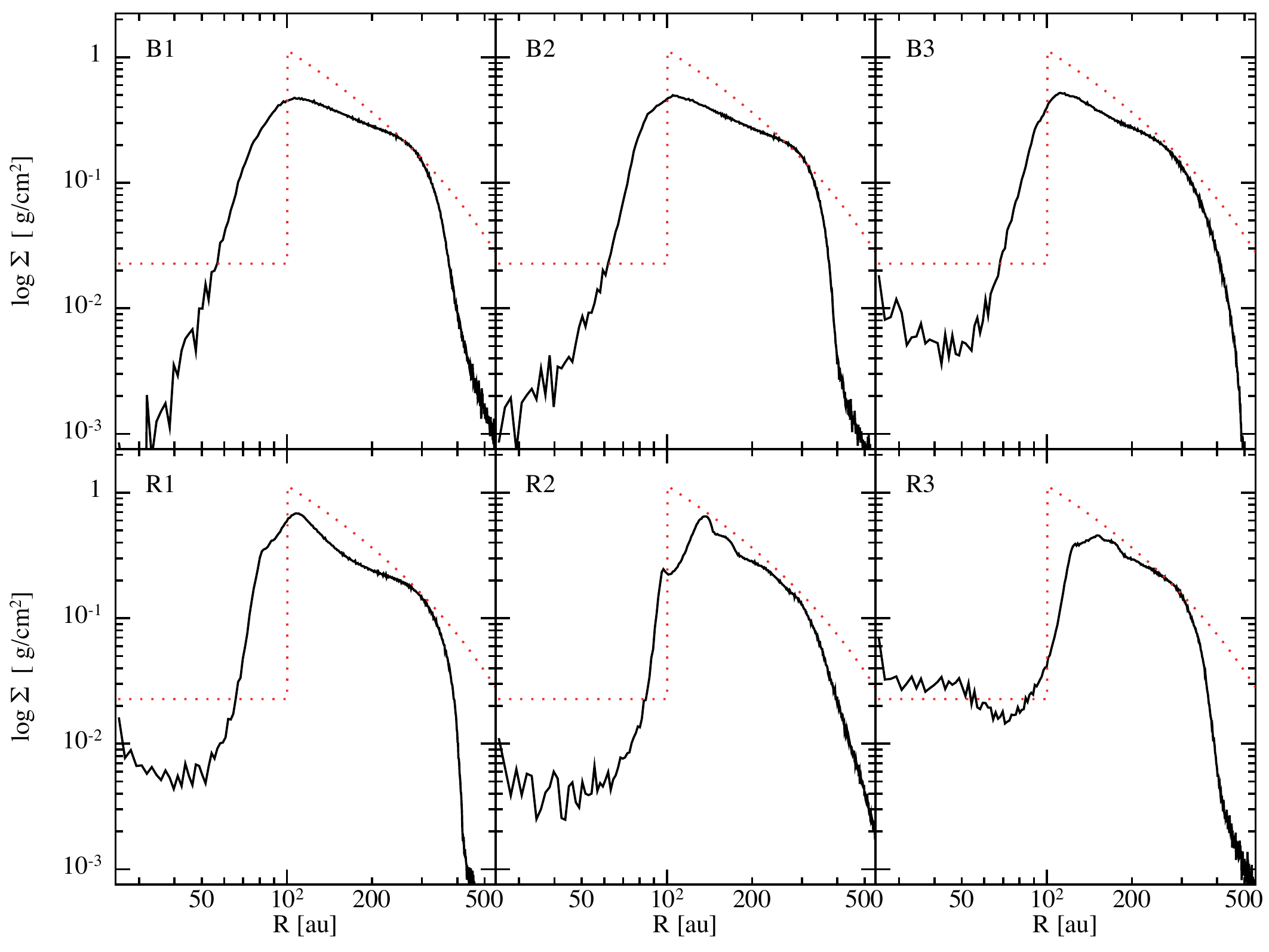}
\caption{Cavity. Surface density as a function of cylindrical radius (in the sky plane) from the $R_{\rm in}=90$ au calculations after 50 binary orbits. Red dotted line shows the model adopted by \citet{perezetal15} from radiative transfer fitting to CO gas. Simulation R2 shows the closest match to the observed cavity size, with $R_{\rm cav}$ within a few percent of the gas cavity size inferred from the CO data. }
\label{fig:cavityrad}
\end{center}
\end{figure*}

\subsubsection{Circumbinary disc}
We performed two sets of calculations, both with a 0.01 M$_\odot$ circumbinary gas disc but with inner radius set initially to either 50 or 90 au. The  $R_{\rm in}=50$ au calculations ensure that clearing of the inner regions is entirely due to tidal effects, while starting with $R_{\rm in}=90$ au avoids transient formation of circumprimary and circumsecondary discs. We set the outer radius to $R_{\rm out} = 350$~au (purely for computational efficiency; the real disc extends to $\sim 600$ au). We assume an initially power law surface density profile $\Sigma \propto R^{-1}$ and model the disc with 10$^6$ SPH particles assuming a total gas mass of 0.01 M$_{\odot}$ --- see \citet{priceetal18} for details of the disc setup used in {\sc Phantom}. We adopt a mean Shakura-Sunyaev disc viscosity $\alpha_{\rm SS} \approx 0.005$ by setting a fixed artificial viscosity parameter $\alpha_{\rm AV} = 0.25$ in the code and using the `disc viscosity' flag (see \citealt{lodatoprice10}). Since the artificial viscosity is explicitly specified in SPH, the value of $\alpha_{\rm SS}$ is directly related to $\alpha_{\rm AV}$ according to
\begin{equation}
\alpha_{\rm SS} = \frac{1}{10} \alpha_{\rm AV}  \frac{\langle h \rangle}{H},
\end{equation}
where the ratio $\langle h \rangle$ is the mean resolution length at a given radius and $H$ is the disc scale height. There are no additional sources of numerical diffusion in SPH (see \citealt{lodatoprice10} for detailed calibration and tests). For our chosen resolution the mean $h/H$ in the outer disc is 0.2 (i.e., 5 resolution lengths per scale height), hence the values of $\alpha_{\rm SS}$ and $\alpha_{\rm AV}$ given above.

 We prescribe temperature as a function of (spherical) radius according to $T= 28 {\rm K} (r/r_{\rm in})^{-0.3}$, consistent with the disc model fit by \citet{casassusetal15} to the observational temperature profile (we use spherical rather than cylindrical radius to avoid confusion when the disc is warped or inclined; see \citealt{lodatoprice10}). This corresponds to $H/R = 0.066$ at $R=R_{\rm in}$ and $H/R = 0.11$ at $R=R_{\rm out}$.  More recent observations \citep{mutoetal15,boehleretal17} found higher temperatures which may suggest a thicker disc. Further modelling of the temperature changes in the disc due to the companion is beyond the scope of this paper but may be important \citep{verhoeffetal11}.
 
 The disc was oriented by $i = 160^\circ$ with respect to the $z=0$ plane, rotated about a Position Angle of $-20^\circ$. That is, our disc is inclined by $20^\circ$ to the line of sight but with the disc rotating clockwise on the sky, consistent with (e.g.) \citet{casassusetal15}.

\begin{figure*}
\begin{center}
\includegraphics[width=\textwidth]{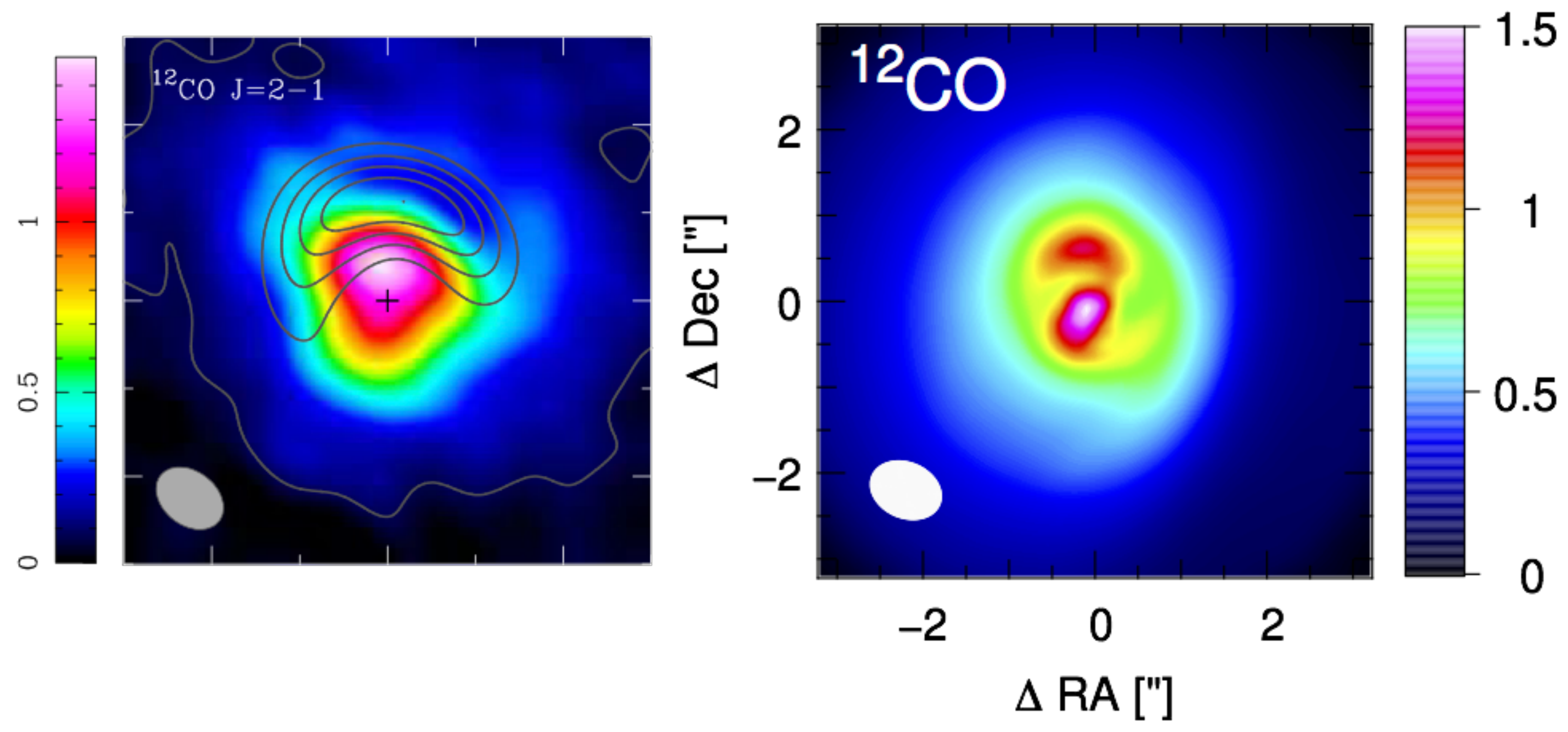}
\vspace{-0.5cm}
\caption{Gas inside the cavity. Predicted $^{12}$CO J=2--1 from simulation R2 (right), compared to ALMA observations (left; credit: Figure 1 of \citet{perezetal15} \copyright AAS, reproduced with permission). Sufficient gas remains inside the cavity to produce optically thick $^{12}$CO emission, as observed.}
\label{fig:12co}
\end{center}
\end{figure*}

\subsection{Radiative transfer}
To perform a direct comparison with observations of HD142527 we post-processed a subset of our simulations using the Monte Carlo radiative transfer code {\sc mcfost} \citep{pinteetal06,pinteetal09}. {\sc mcfost} is particularly suited to post-processing data from SPH codes because it uses a Voronoi tesselation where each cell corresponds to the position of an SPH particle. Properties such as density, temperature and velocity may then be mapped directly from the particles to the radiative transfer grid without interpolation.

To irradiate the disc we used the two sink particles as isotropic sources with stellar spectral models appropriate to their mass. Specifically, we adopt 3Myr
 isochrones from \citet{sdf00}, with T$_{\rm eff} = 4800\,$K and a luminosity of 2.7\,L$_\odot$ for the primary and T$_{\rm eff} = 3700\,$K and L =
 0.22\,L$_\odot$  for the secondary. We assume astronomical silicates \citep{draine03a}, with a grain size distribution ranging from 0.03$\,\mu$m to 1\,mm, and a slope of $-3.5$. Optical properties are calculated using Mie theory. We then used {\sc mcfost} to predict line emission from our gas-only simulations for the isotopologues of carbon monoxide including $^{12}$CO, $^{13}$CO and C$^{18}$O as well as HCO$^+$. We assume a uniform abundance of $5 \times 10^{-5}$, $7 \times 10^{-7}$, $2 \times 10^{-7}$ and $10^{-9}$ for $^{12}$CO, $^{13}$CO, C$^{18}$O, and HCO$^+$, respectively. We assume the gas is in local thermodynamic equilibrium with T$_{\rm gas}$ = T$_{\rm dust}$. We predicted only the line emission, not the continuum, from our gas-only simulations under the assumption that the dust follows the gas. This is a good assumption for grain sizes $\lesssim$ 10 $\micron$ but for larger grains this assumption no longer holds (see Section~\ref{sec:dust}).


\begin{figure*}
\begin{center}
\includegraphics[width=\textwidth]{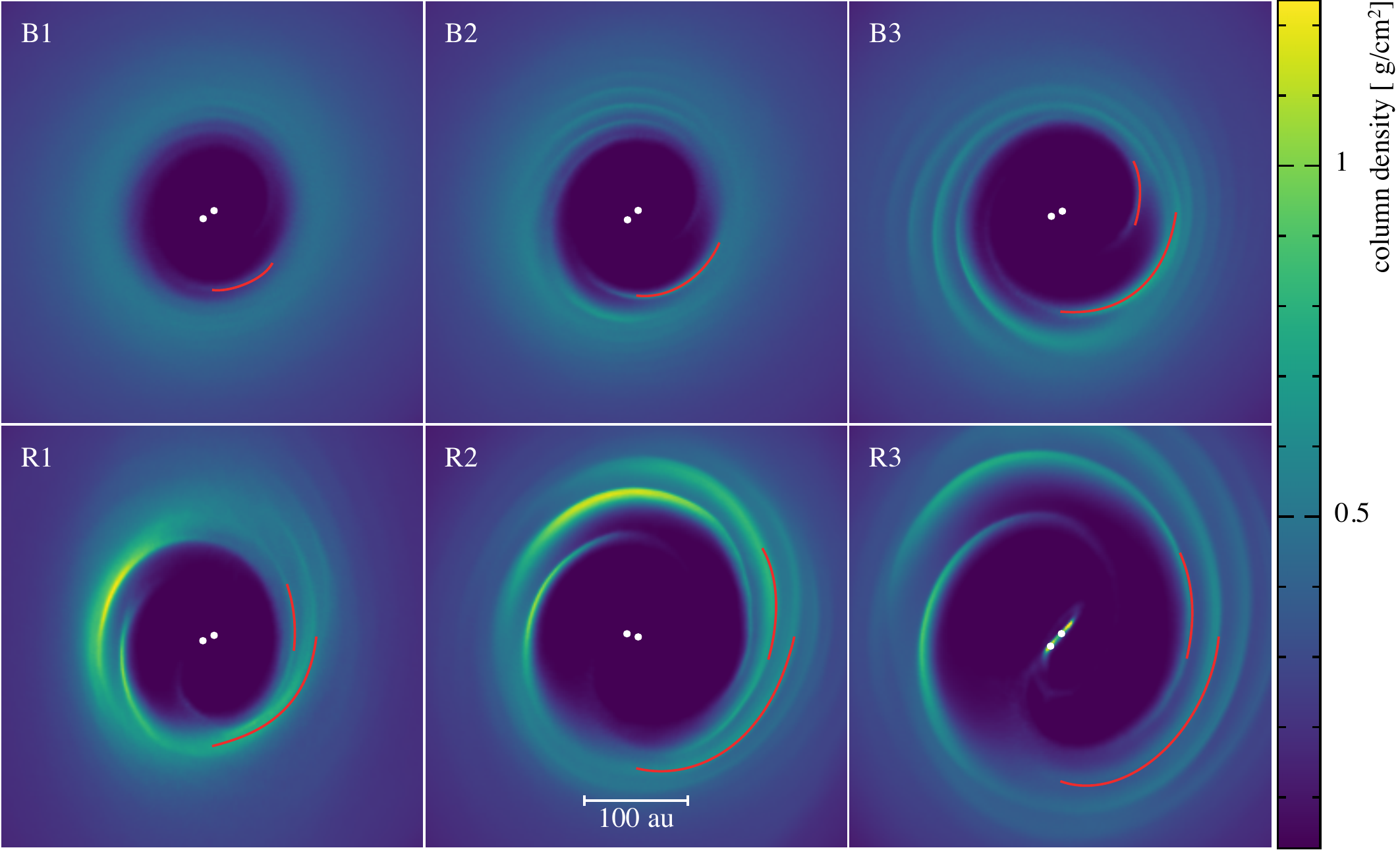}
\vspace{-0.4cm}
\caption{Spirals. Gas surface densities after 50 binary orbital periods in calculations with initial $R_{\rm in} = 90$ au, showing the spiral arms. As in Figure~\ref{fig:cavity}, top row shows the `blue' orbits (binary just past periastron) while bottom row corresponds to the `red' orbits (binary approaching periastron) from \citetalias{lacouretal16}. Comparison with the observed spiral structure (left panel of Figure~\ref{fig:circumprimary}) favours the latter. Comparison with Fig.~\ref{fig:cavity} shows the cavity size is independent of the initial $R_{\rm in}$. Red lines show polynomial fits to `inner' and `outer' spiral arms tracing a position angle range similar to the scattered light spirals (Fig.~\ref{fig:circumprimary}).} 
\label{fig:spirals}
\end{center}
\end{figure*}

\section{Results}
\label{sec:results}

\begin{table}
\begin{minipage}{\columnwidth}
\begin{tabular}{lp{1.45cm}|p{1.2cm}|p{1.15cm}|p{1.15cm}|p{0.6cm}|}
& M ($< 90$ au) $(\times 10^{-3} {\rm M}_\odot)$ & Pitch ang. (PA $265^\circ$) & Pitch ang. (1st half) & Pitch ang. (2nd half) & SW Fork? \\
\hline
Obs & $1.7 \pm 0.6$ & 4.8$^\circ$& -3.6$^\circ$ & 9.9$^\circ$ & Yes \\
B1 & 2.0 & 1.5$^\circ$& 3.1$^\circ$ & 0.0$^\circ$ & No \\
B2 & 1.9 & 1.3$^\circ$& 3.2$^\circ$ & 2.5$^\circ$ & No \\
B3 & 1.6 & 5.1$^\circ$& 7.1$^\circ$ & 3.2$^\circ$ & No \\
R1 & 2.1 & -1.8$^\circ$& -4.0$^\circ$ & 0.6$^\circ$ & Yes \\
R2 & 1.5 & 6.0$^\circ$& 6.4$^\circ$ & 5.7$^\circ$ & No \\
R3 & 1.4 & 3.6$^\circ$& 6.8$^\circ$ & 0.4$^\circ$ & No \\
\end{tabular}
\caption{Gas mass interior to 90 au (first column); pitch angles along the outer spiral arm at position angle of $265^\circ$ (second column) and within the first and second half of the spiral in position angle (third and fourth columns); comparing our simulations to the observations (top row) after 50 orbits of the binary. The final column indicates whether or not the spiral arms show a bifurcation or `fork' towards the south-west. All simulations obtain residual mass interior to the cavity consistent with the observations. Simulations B3, R2 and R3 show pitch angles most consistent with the data.}
\label{tab:spirals}
\end{minipage}
\end{table}

\subsection{Cavity}
Figure~\ref{fig:cavity} shows the column density view of the six calculations corresponding to the orbits listed in Table~\ref{tab:spirals}, shown after 20 binary orbital periods with initial $R_{\rm in} = 50$ au. Transient circumprimary discs are present in all calculations except B1. For both `blue' and `red' sets of orbits the cavity size scales with semi-major axis (left to right; \citealt{artymowiczlubow94}). Cavity size also increases with eccentricity due to the increased apocentre separation (but decreases with inclination; \citealt{mirandalai15}). Comparison with the cavity size seen in scattered light, we can exclude binary orbits with $a \gtrsim 50$~au. That with plausible orbits for the observed binary we can already produce a cavity size \emph{too big} solves one of the major mysteries of HD142527 --- how a binary with a projected separation of 13~au can produce a 100~au cavity. 

 Figure~\ref{fig:cavityrad} shows the surface density binned as a function of cylindrical radius in the sky plane ($z=0$). The red dotted line shows the parameterised model adopted by \citet{perezetal15} from fits to the CO emission (specifically, we plot equation 6 from their paper, with parameters taken from their best fit model corrected for the GAIA distance, giving $R_{\rm cav} = 100$ au). Simulation R2 produces the closest match to the data, with $R_{\rm cav}$ within a few percent of the observational fit. The remarkable agreement demonstrates that orbits consistent with those inferred by \citetalias{lacouretal16} indeed produce cavities of the correct size.


\subsection{Gas inside the cavity}
 The drop in surface density seen in Figure~\ref{fig:cavityrad} indicates that the cavity interior is not completely devoid of gas. To quantify this, Table~\ref{tab:spirals} lists the mass interior to 90 au in each of our simulations, compared to the observational measurement from \citet{perezetal15}. Regardless of our choice of binary orbit the residual mass inside the cavity agrees with the observational estimate to within the error bars. For example, we measure a residual gas mass of $1.5 \times 10^{-3}$~M$_\odot$ within 90 au in simulation R2 after 50 orbits, within 11\% of the $(1.7 \pm 0.6) \times 10^{-3}$ M$_\odot$ measured by \citet{perezetal15}. 
 
 Figure~\ref{fig:12co} shows the predicted $^{12}$CO J=2--1 emission from simulation R2 (right panel), convolved to a beam size consistent with the \citet{perezetal15} Cycle 0 ALMA data (shown in left panel). We find remarkable agreement with the observed $^{12}$CO emission from inside the cavity. In particular there is sufficient gas inside the cavity such that the $^{12}$CO is optically thick, as observed. The binary scenario thus naturally explains the residual gas found within the cavity around HD142527.

\begin{figure*}
\begin{center}
\includegraphics[width=0.4\textwidth]{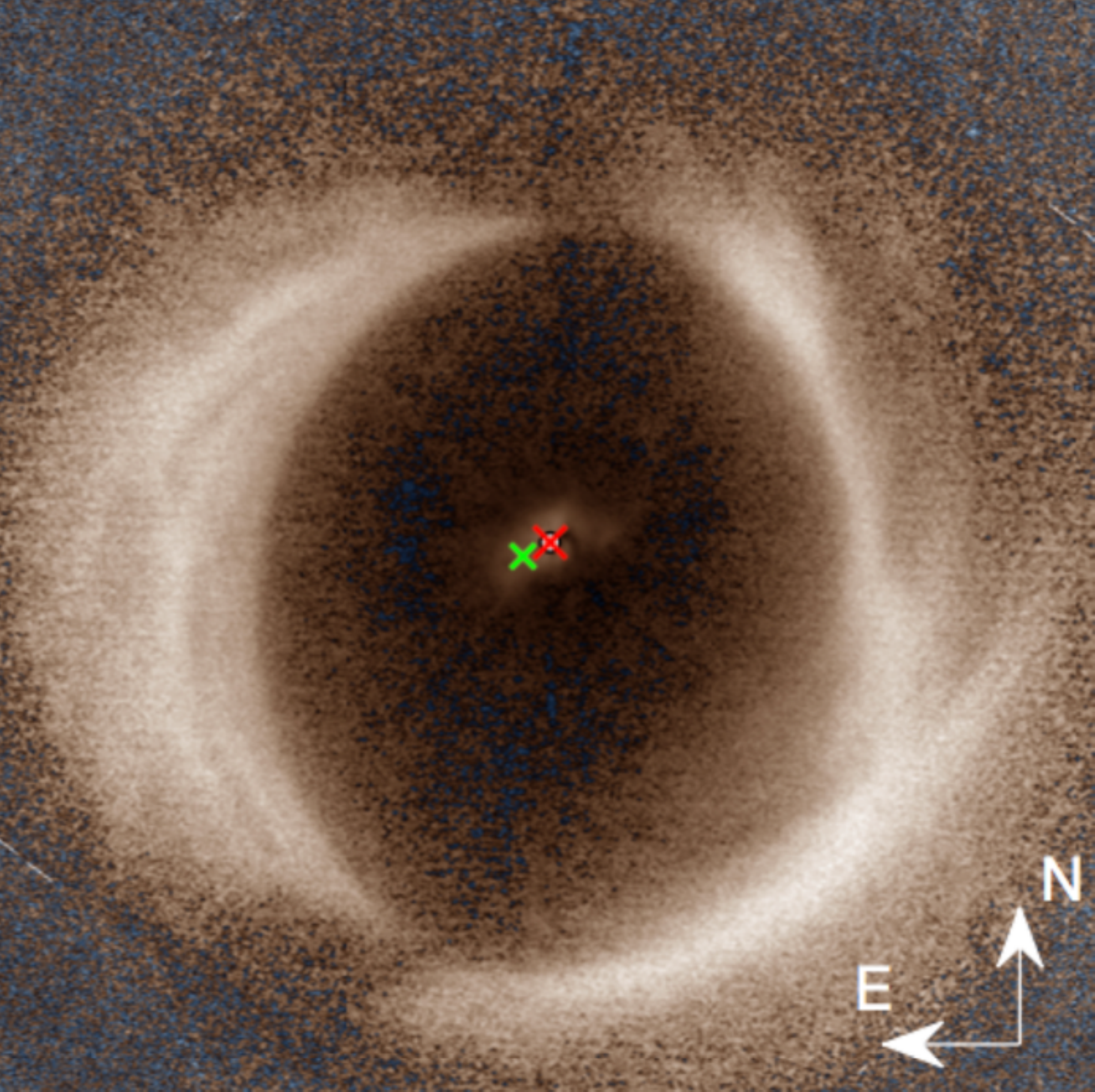}
\includegraphics[width=0.4\textwidth]{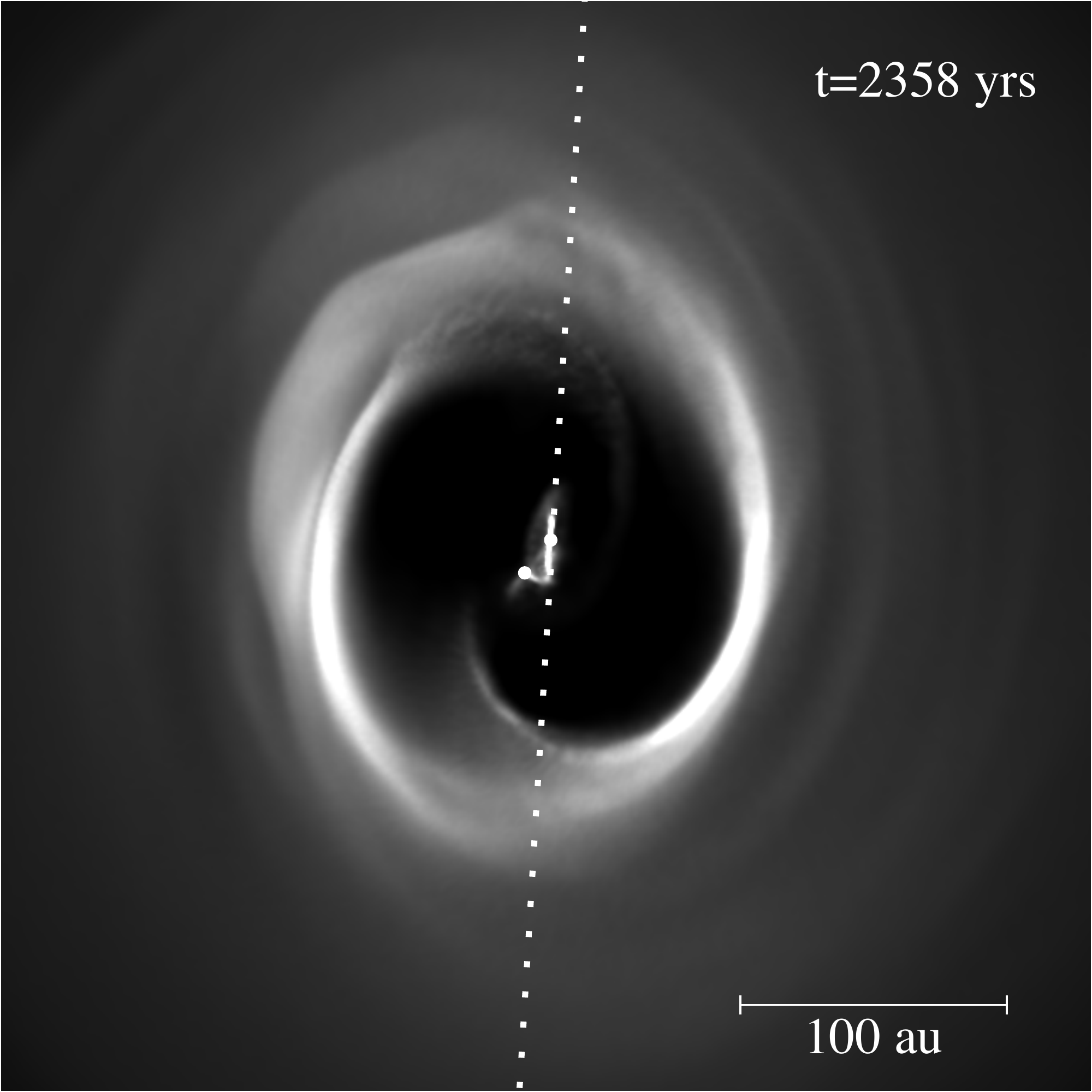}
\caption{Shadows. Column density in the R1 orbit simulated with initial $R_{\rm in}=50$au after 20 orbits at the observed orbital phase (right), showing the orientation of the (transient) circumprimary disc, compared to the scattered light (600-900nm) ZIMPOL polarisation image (left; taken from Figure 1 of \citet{avenhausetal17} \copyright AAS, reproduced with permission). Dotted line indicates the expected shadow from our simulated inner disc (right), which lies close to the orientation of the observed shadows (left).}
\label{fig:circumprimary}
\end{center}
\end{figure*}

\subsection{Spirals}
 Figure~\ref{fig:spirals} shows the column density from the $R_{\rm in}=90$ au calculations after 50 binary orbital periods. In every case three or more prominent spiral arms are seen around the cavity. The companion thus already explains why spiral arms are present around the cavity without needing to invoke gravitational instability or other physics. Comparing the $R_{\rm in}=50$~au calculations shown in Figure~\ref{fig:cavity} shows that both calculations produce the same cavity sizes, with the cavity size approximately constant after $\gtrsim 10$ orbits of the central binary.
 
 Comparing our results with scattered light \citep{avenhausetal14,avenhausetal17}, 2 micron \citep{casassusetal12,casassusetal13} images of HD142527 shows that the blue orbits (top row) tend to produce spirals inconsistent with the observations. For a more quantitative comparison, we fit the spiral structure with a polynomial of the form $r(\theta) = \sum_{i=0}^{4} a_i \theta^i$. Aside from the formula used to fit the spirals, the fitting procedure is otherwise identical to that described in \citet{christiaensetal14}. That is, we trace pixels along the maxima in column density and fit a polynomial to the resulting points using a least squares fit. Using this procedure we fit `inner' and `outer' spiral arms spanning the position angle ranges $[260^\circ,305^\circ]$ and $[180^\circ,270^\circ]$, respectively, corresponding to the two main spirals seen to the south west of the cavity in the scattered light image. Red lines in Figure~\ref{fig:spirals} show the corresponding fits. For orbits B1 and B2 no inner spiral could be fitted, so we only show the outer arm. 
 
 Table~\ref{tab:spirals} lists the resulting pitch angles for the outer spiral measured at a position angle (PA) of $265^\circ$, and with the first and second half of the spiral (in position angle), corresponding to PA ranges $[180^\circ, 225^\circ]$ (third column) and $[225^\circ, 270^\circ]$ (fourth column). Orbits B1 and B2 in particular show pitch angles too small and little or no asymmetry in the gas distribution around a cavity which is close to circular. Orbit B3 shows promising spirals to the north-east of the cavity, but the spiral arms to the south-west show a series of almost-circular tightly wrapped arms not seen in the observations. 
 
 The red orbits, by contrast, produce spiral structure and asymmetry in both the cavity and the gas distribution more similar to what is observed (bottom row). In particular, orbit R1 is the only simulation to show a bifurcation in the spiral arms to the south-west of the cavity, as seen in the scattered light image (see Table~\ref{tab:spirals}). Orbit R2 is closest to the observed cavity size (Fig.~\ref{fig:cavityrad}) and develops an eccentric cavity similar to those found in previous circumbinary disc simulations \citep[e.g.][]{farrisetal14,dorazioetal16,rlp16} and used by \citet{ragusaetal17} to explain dust horseshoes. The spiral arms appear fixed in relation to the binary orbit, but superimposed on this is a precessing overdensity which precesses on a timescale of 2--3 orbital periods, as found by \citet*{dcd15}. Orbit R3 shows the most open spiral arms (Table~\ref{tab:spirals}) but the overdensity around the cavity is less prominent, making it less promising for the observed dust structures. The cavity size is also too big (Fig.~\ref{fig:cavityrad}).

\begin{figure*}
\begin{center}
\includegraphics[width=0.4\textwidth]{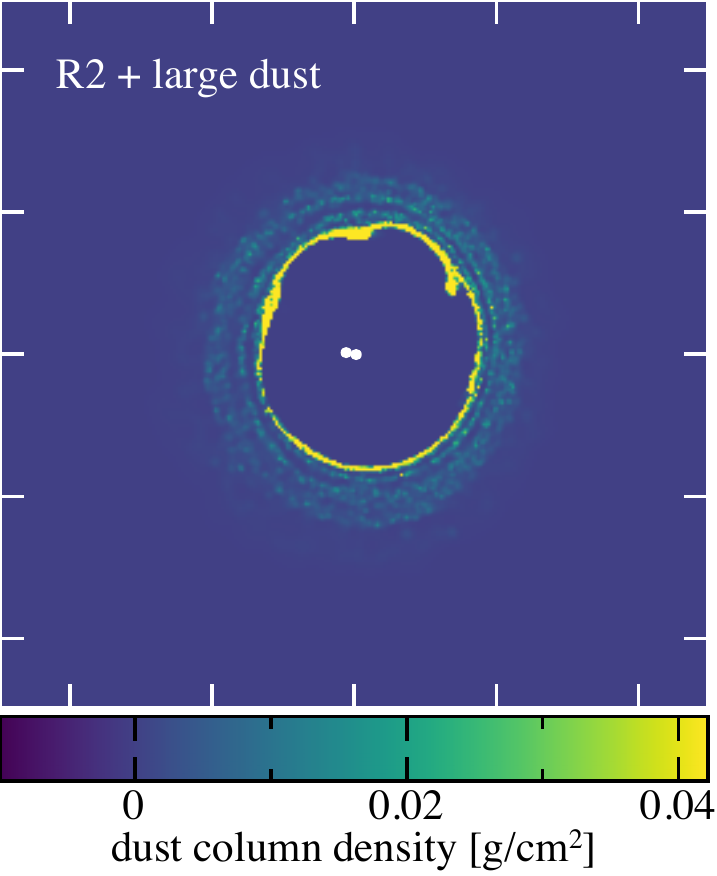}
\hspace{0.4cm}
\includegraphics[width=0.4\textwidth]{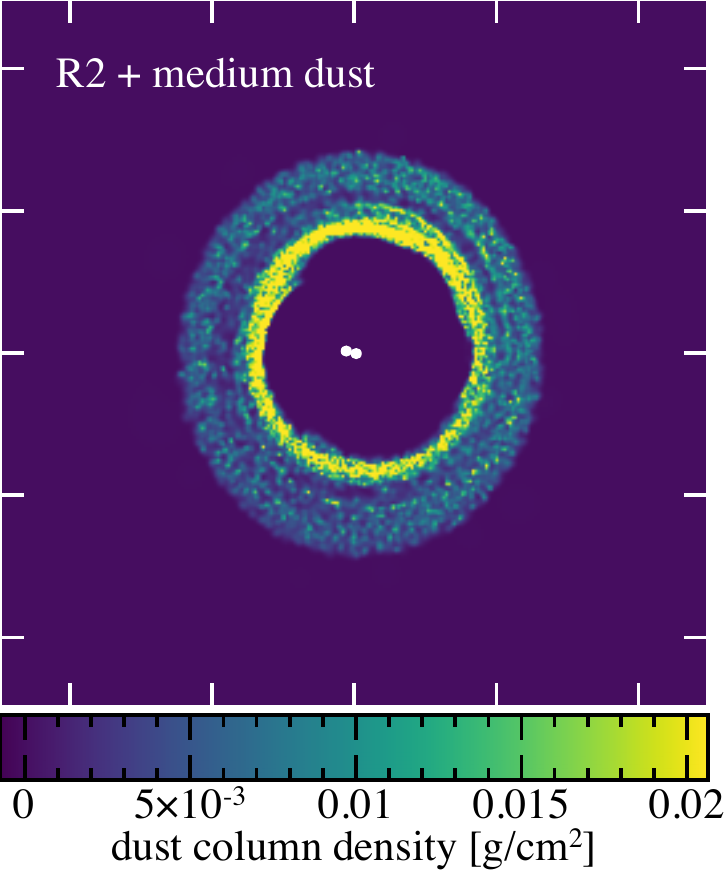}
\caption{Dust. Dust column density in dust-gas simulations using orbit R2 with mm grains (left panel) and 100$\micron$-sized grains (right panel). Our `large' (mm) grains are close to Stokes number of unity, and hence quickly migrate to form a thin ring at the cavity edge. Decreasing the grain size by a factor of ten (right panel) produces a more radially extended dust structure.} 
\label{fig:dust}
\end{center}
\end{figure*}

 \begin{figure*}
\begin{center}
\includegraphics[width=0.9\textwidth]{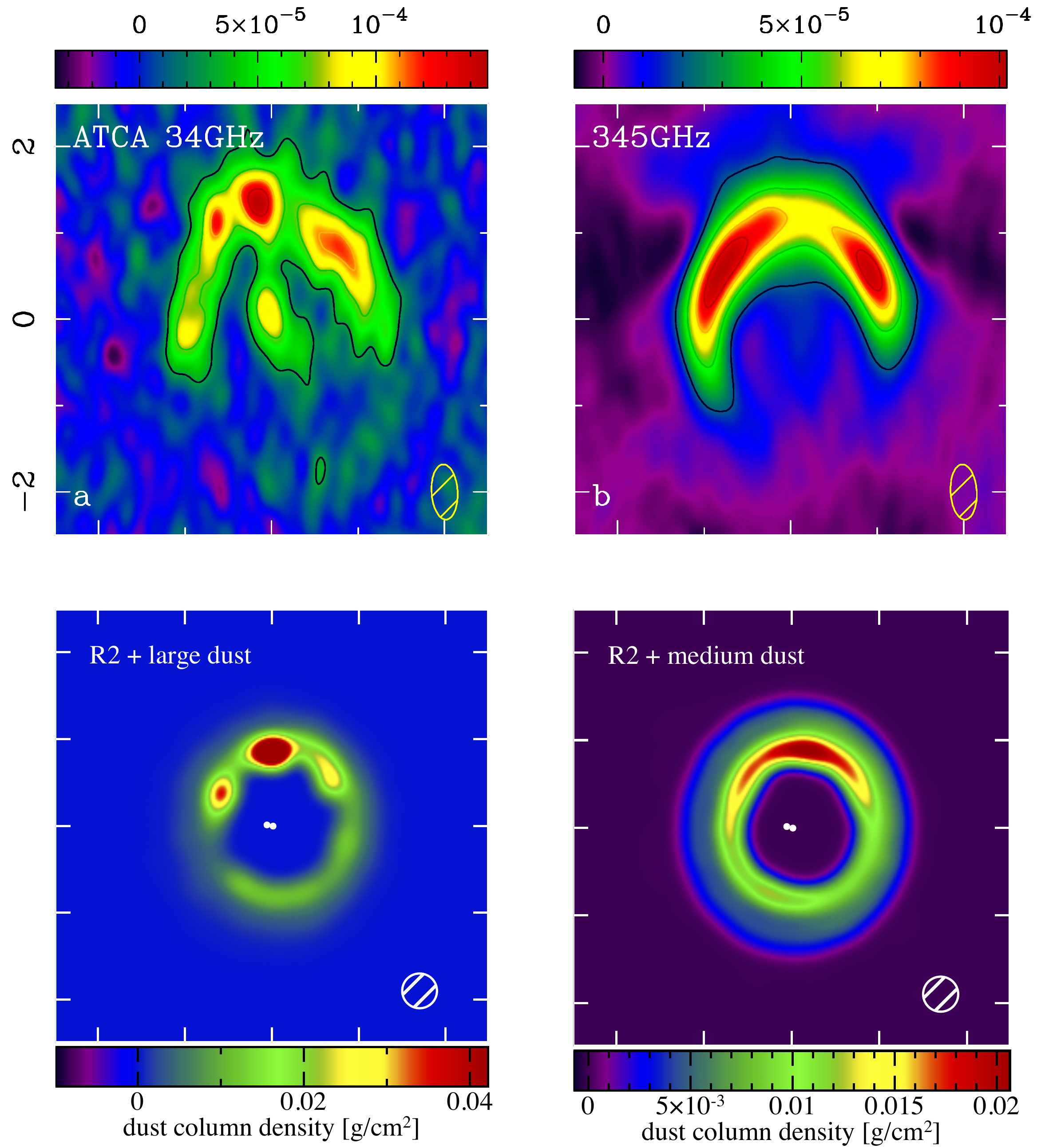}
\caption{Horseshoe. Dust column density comparing observations (top) to our simulations (bottom). We show orbit R2 with grain sizes of 1 mm (`large') and 100 $\micron$ (`medium') (bottom panels, left and right respectively) smeared to a beam size of 20 au (white circle). The larger grains (bottom left) show dust column densities consistent with the ATCA observations at cm wavelengths while the smaller grains (bottom right) produce an overdense `horseshoe' in the dust surface density at the correct position angle to explain the ALMA observations (top right). Top row credit: Figure 6 of \citet{casassusetal15a} \copyright AAS, reproduced with permission.} 
\label{fig:horseshoe}
\end{center}
\end{figure*}

\subsection{Shadows}
Figure~\ref{fig:cavity} shows the formation of transient circumprimary discs in the $R_{\rm in}=50$ au calculations during the first 20 orbits, caused by the clearing of the inner disc material. The orientation of these inner discs are highly sensitive to the orbit of the companion. For example, the circumprimary disc in calculation B3 is formed with major axis aligned east-west in our images (i.e. horizontal in Figure~\ref{fig:cavity}) , while orbits R1 and R2 produce a disc aligned north-south (i.e. vertical in Figure~\ref{fig:cavity}) --- a second piece of evidence favouring the red orbit family. Caution is required, however, since the inner disc precesses with time, though on a timescale longer than our simulations ($\sim 0.5$Myr; e.g. \citealt{owenlai17}). Furthermore, in our $R_{\rm in}=90$ au calculations we find rotationally supported circumprimary discs only with the R3 orbit at this resolution (Figure~\ref{fig:spirals}; this mainly indicates that the disc mass in other calculations is too low or that the numerical viscosity drains the disc too fast at this resolution; not that circumprimary discs do not exist).
 
Figure~\ref{fig:circumprimary} shows the resultant inner disc structure after 20 orbits (left), shown alongside the scattered light image taken from \citet{avenhausetal17} (left). Despite the remaining uncertainty in the orbital dynamics the position angle of the expected shadow is consistent with the northern shadow seen in the scattered light image, and within 10$^\circ$ of the southern shadow. A difference of 10$^\circ$ is not significant --- shadows do not exactly fall in the projected plane of the inner disc due to the vertical extension of the discs \citep{minetal17}. We do not imply that this is the \emph{only} possible orbital configuration which can explain the shadows --- nor even the most probable --- merely that it is possible to produce a satisfactory orientation of the inner disc to produce the correct shadow from our calculations. We also demonstrate that the orbital dynamics of the companion naturally produces a circumprimary disc with an orientation and size consistent with the structure invoked by \citet{mpc15} to explain the observed shadowing.


\begin{figure*}
\begin{center}
\includegraphics[width=\textwidth]{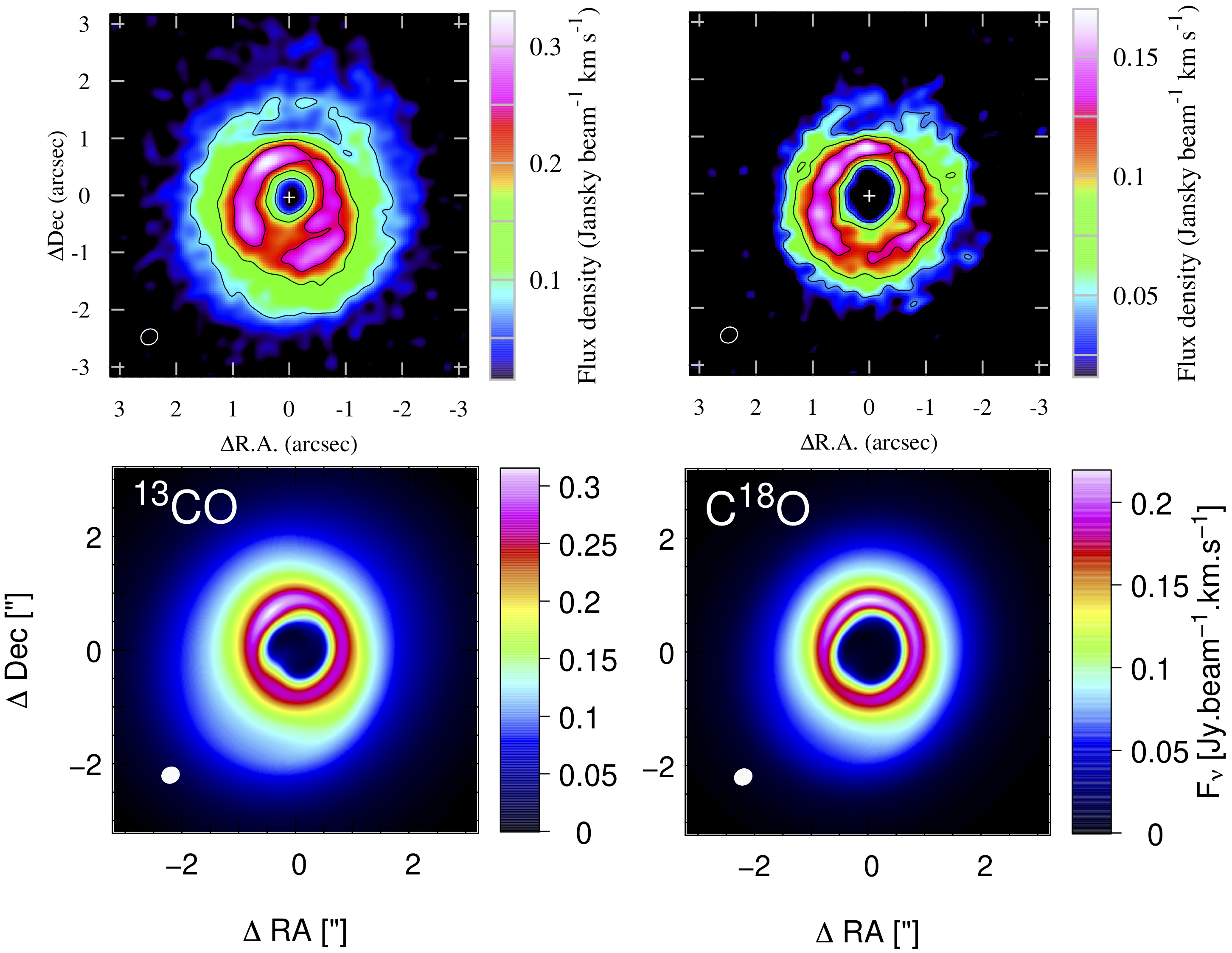}
\caption{Gas. Predicted $^{13}$CO J=3--2 and C$^{18}$O J=3--2 emission (moment 0) from simulation R2 after 50 orbits (bottom two panels), compared to \citet{boehleretal17} ALMA data (top two panels; credit: Figure 1 of \citet{boehleretal17} \copyright AAS, reproduced with permission). The ring-like feature seen in both spectral lines is reproduced in our simulations. The `broken ring' effect seen in the observations is not reproduced in our models because we do not account for the temperature dip caused by the shadows from the circumprimary disc.}
\label{fig:13co}
\end{center}
\end{figure*}

\subsection{Dust horseshoe}
\label{sec:dust}

\citet{ragusaetal17} found that dust horseshoes similar to those observed by ALMA could be naturally produced by eccentric cavities in gas and dust around binary stars. However, that paper assumed tightly coupled dust grains such that the dust structures merely reflected those in the gas. Whether or not this is the case for the mm continuum emission in HD142527 (see discussion in \citealt{casassusetal15a}) depends on the Stokes number --- the ratio of the dust stopping time to the disc orbital period. 

Assuming subsonic Epstein drag the Stokes number depends only on the gas surface density, according to \citep{dipierroetal15}
\begin{equation}
S_{\rm t} = 1 \bigg( \frac{\Sigma}{0.2 \textrm{ g cm}^{-2}} \bigg)^{-1} \bigg( \frac{\rho_{\rm grain}}{3 \textrm{g cm}^{-3}} \bigg) \bigg( \frac{s_{\rm grain}}{1 \textrm{ mm}} \bigg),
\end{equation}
where $\rho_{\rm grain}$ is the intrinsic grain density and $s_{\rm grain}$ is the grain size. Modelling dust-gas dynamics in discs is usually uncertain because the gas surface density is poorly constrained, being measured only from multiplying the dust continuum emission by a factor (typically 100). For HD142527 we are already confident in our assumed surface density profile because of our match to the measured gas mass inside the cavity and to the surface density profile (Fig.~\ref{fig:cavityrad}). However, there remains uncertainty in the assumed CO-to-H$_2$ conversion.

Our assumed gas disc mass of 0.01 M$_{\odot}$ corresponds to $\Sigma$ = 0.6 g cm$^{-2}$ at $R = R_{\rm in}$, giving a Stokes number for mm grains of $\approx$ 0.3 at 90 au and $\approx 1$ at 300 au (see Fig.~\ref{fig:cavityrad}). Thus we expect decoupling of grains in the outer disc, since $S_{\rm t}=1$ corresponds to maximal radial drift \citep{weidenschilling77}.

 Since orbit R2 produces a cavity close to the observed size with a prominent asymmetry seen in the gas in the position angle similar to the observed mm horseshoe, we computed two additional simulations using grain sizes of $1$ mm and 100 $\micron$, respectively. We set up dust disc initially between $R_{\rm in}=120$ and $R_{\rm out} = 250$ au using $2.5 \times 10^4$ dust particles, with an (arbitrary) initial dust-to-gas ratio of 0.01 and properties otherwise reflecting the gas disc (composed of the usual $10^6$ SPH particles). The smaller dust disc is merely to avoid numerical problems during the initial disc response to the binary. This also ensures that any dust migration to the cavity edge occurs naturally rather than as a result of the initial conditions.
 
  Figure~\ref{fig:dust} shows the dust column density in these two simulations (left and right, respectively), shown after 62 and 63 binary orbits, respectively, such that the orbital phase of both the binary and the dust structures are consistent with the observations.  We find that the dust in both simulations drifts radially, concentrating at the cavity edge within a few tens of orbits. The larger grains, with $S_{\rm t} \sim 1$, form a thin ring around the cavity edge (left panel). We also observe dust grains collecting into azimuthally distinct structures, trapped by the pressure bumps at the locations where spiral arms in the gas cross the dust ring. Although these grains are nominally `mm-sized' in our simulations, the resulting dust structures appear more similar to what is seen at cm wavelengths. Figure~\ref{fig:horseshoe} shows a direct comparison with the ATCA 34 GHz image from \citet{casassusetal15a} (top left shows the ATCA image; bottom left shows our simulation). To make this comparison we simply convolved our dust column density image to a beam size of 20 au (0.12"). In hindsight this is not surprising, since the peak emission in wavelength is roughtly $2\pi$ times the grain size. 
  
  Using grains ten times smaller produces slower migration to the cavity edge and, as a result, a more radially extended dust distribution (right panel of Figure~\ref{fig:dust}). The asymmetry driven by the binary in the gas produces a horseshoe remarkably similar to the observed mm horseshoe. Figure~\ref{fig:horseshoe} makes the direct comparison. The predicted continuum emission is shown in red in Figure~\ref{fig:hco}.
 
 The main disagreement between the simulations and observations concerns the prominent dip seen in the mm-horseshoe at a position angle of $\sim 10^\circ$ (top right). Caution is required in making this comparison since our visualisation shown in Figure~\ref{fig:horseshoe} assumes optically thin dust emission where spectral index variations show that the continuum emission is optically thick \citep{casassusetal15a}. Our models also do not account for any azimuthal variation in temperature. This is interesting, because a temperature decrement is indeed observed at this position angle caused by the shadow from the circumprimary disc. Such a temperature decrement will affect the dust emission and would need to be accounted for in making accurate radiative transfer predictions from our simulations. Moreover, it suggests that thermodynamic effects from the shadow may be important (see \citealt{montesinosetal16}).
  
  Two conclusions stand out: i) decoupled dust dynamics around a binary-carved cavity can naturally explain the observed asymmetries in HD142527 without recourse to vortices or additional companions; and ii) the distinct `blobs' seen in the radio emission may be real and not just artefacts of noisy observations.

\subsection{Gas density contrast}
 \citet{casassusetal15a} note that some of the asymmetry seen in the dust emission is also seen in the gas. To quantify this, Figure~\ref{fig:13co} compares the predicted $^{13}$CO J=3--2 and C$^{18}$O J=3--2 (moment 0) emission maps from simulation R2 (bottom row) to the ALMA observations recently published by \citet{boehleretal17} (top row; see also \citealt{mutoetal15}). For both spectral lines we find a bright, asymmetric ring of emission surrounding the cavity at a radius between 0.5 and 1" from the central source, as seen in the observations. 
 
 The main source of disagreement is that we do not reproduce the two dips in emission at the position angles coincident with the scattered light shadows. This again suggests that the shadowing of the outer disc by the circumprimary disc needs to be accounted for in the CO emission. Our models suggest that the underlying gas density structure is more axisymmetric. If one neglects the dips caused by the shadows, then the contrast in emission around the cavity is similar between our simulations and the observations (i.e. roughly a factor of two).
 
  Again, the close match with the observed line emission does not suggest alternative hypotheses other than the binary are needed.

\begin{figure}
\begin{center}
\includegraphics[width=\columnwidth]{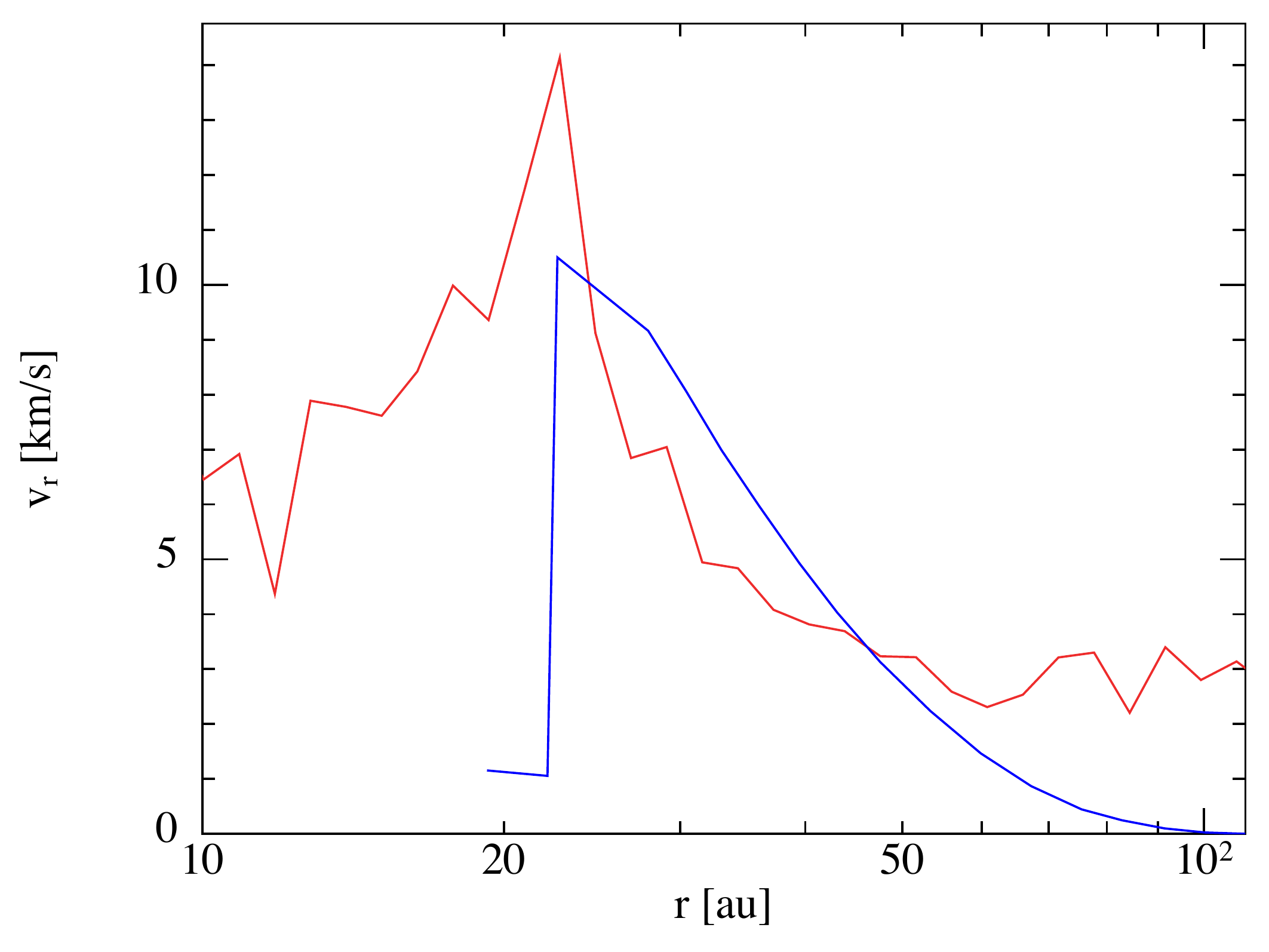}
\caption{Fast radial flows. Maximum inflow velocity on the SPH particles binned as a function of radius, in our model R2 (red), compared to the radial velocity model used to fit the kinematic data (from \citealt{casassusetal15a}; solid blue line). Fast radial flows of order 10 km/s occur naturally in the models caused by the streamers which penetrate the cavity.}
\label{fig:vr}
\end{center}
\end{figure}

\begin{figure*}
\begin{center}
\includegraphics[width=\textwidth]{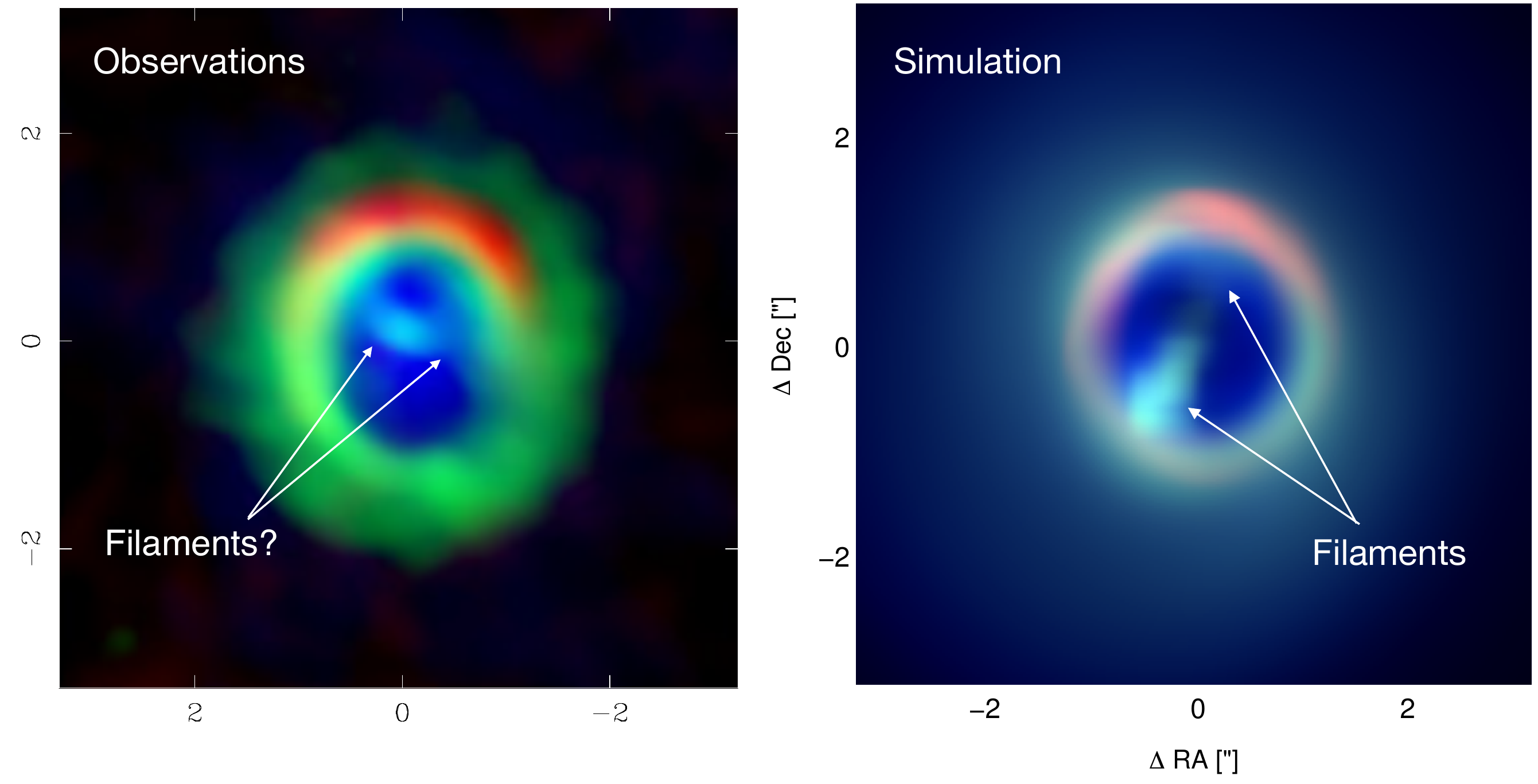}
\caption{Streamers. Predicted HCO+ emission from simulation R2 (right; HCO+ in green), compared to the corresponding Cycle 0 image (left; credit: Figure 8 of \citet{casassusetal15} \copyright AAS, reproduced with permission). Both images show the mm dust horseshoe in red with HCO+ emission shown in green, with $^{12}$CO emission in blue. A filament can be seen crossing the cavity in our simulations, similar to what is observed (albeit with a different position angle, indicating our orbit is not the correct one). We thus identify this feature with the streams of material feeding the primary across the circumbinary cavity.}
\label{fig:hco}
\end{center}
\end{figure*}

\subsection{Fast radial flows}
 Can the streamers seen in Figure~\ref{fig:cavity} explain the fast radial flows? Figure~\ref{fig:vr} shows the radial velocity of the SPH particles, binned as a function of radial distance from the centre of mass (red line), compared to the model used to fit the kinematic data by \citet{casassusetal15a} (blue line). Inflow speeds can be seen to reach 10 km/s at a distance of 20--30 au, consistent with the `fast radial flows' needed to fit the kinematic data. This suggests that these flows indeed originate in the streams of material that feed the inner disc.

\subsection{Gap-crossing filaments}
 One of the great mysteries in HD142527 concerns the origin of the `filaments' of gas seen across the cavity in HCO+ emission by \citet{casassusetal13}. These were seen only in HCO+ emission in a particular range of velocity channels. Figure~\ref{fig:hco} compares the predicted HCO+ J=4--3 emission from our simulations (right panel) to the corresponding figure from \citet{casassusetal15} (left panel). As in the observational figure, we show our predicted continuum map for the mm grains (extrapolated from our "medium grains" dust simulation) in red, with the predicted HCO+ emission in selected velocity channels in green and the predicted $^{12}CO$ emission in blue.
 
  Our predicted HCO+ emission shows a thin, faint filament crossing the cavity (right panel), remarkably similar to what is observed (left panel). This suggests that this feature is indeed of physical origin, originating in the streams of material that feed the circumprimary disc across the cavity. This serves as further confirmation that the dynamical interaction with the binary companion is the source of almost all of the mysterious features present in HD142527. It also suggests that HCO+ should be more widely employed to detect intra-cavity flows in circumbinary discs. The main discrepancy our comparison with observations in Figure~\ref{fig:hco} is that the position angle of the stream differs from our simulation. This mainly indicates that orbit R2 is not the true orbit.

\begin{figure*}
\begin{center}
\includegraphics[width=0.92\textwidth]{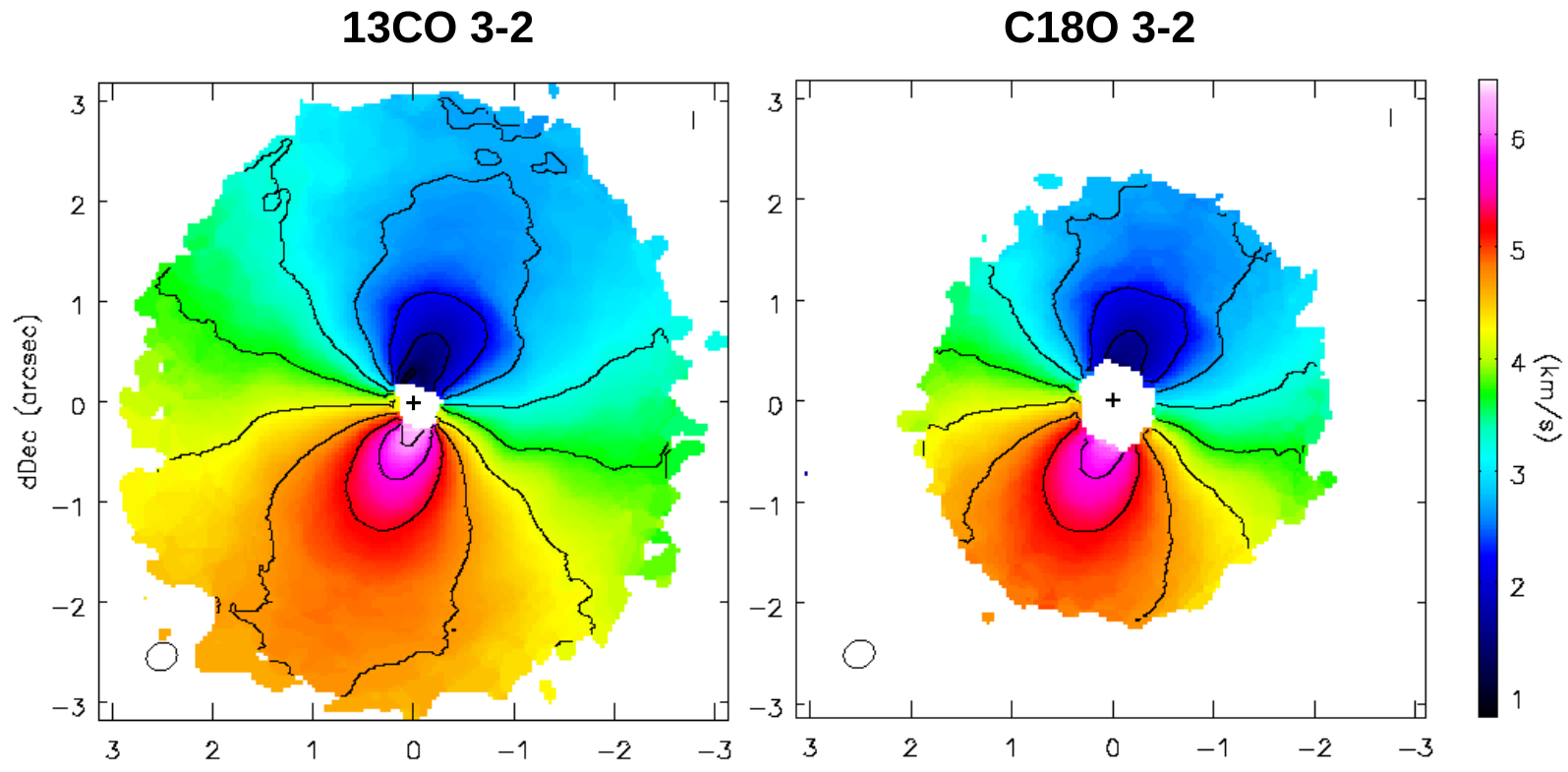}
\includegraphics[width=\textwidth]{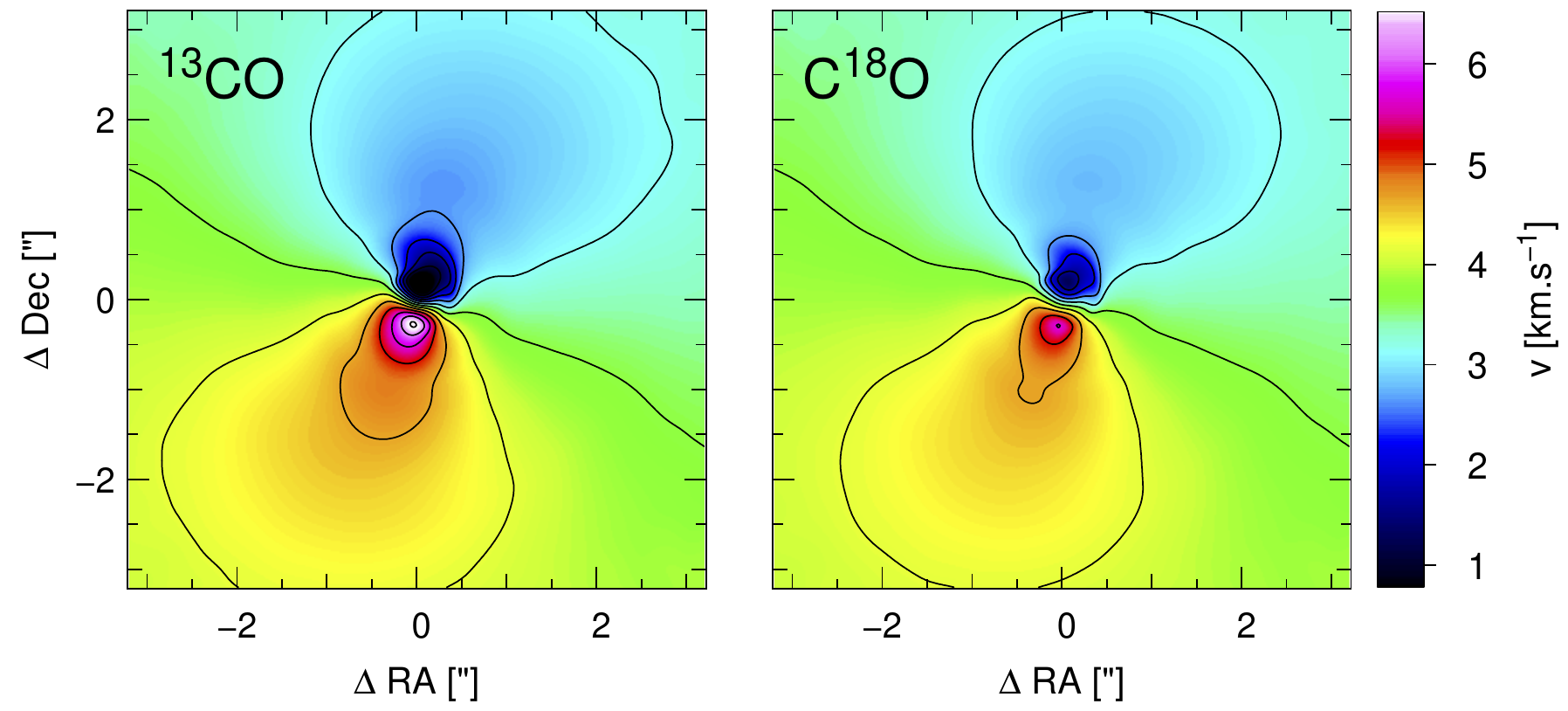}
\caption{Kinematics. Predicted moment 1 emission maps in $^{13}$CO 3--2 and C$^{18}$O from our simulation R2 (bottom left and right, respectively) compared to the observations with ALMA (top; credit: \citet{boehleretal17} \copyright AAS, reproduced with permission).}
\label{fig:m1}
\end{center}
\end{figure*}

\subsection{Kinematics}
Finally, Figure~\ref{fig:m1} compares the predicted moment 1 maps from simulation R2 in $^{13}$CO and C$^{18}$O emission (bottom row) with the equivalent maps taken from \citet{boehleretal17} using ALMA data (top; see also \citealt{mutoetal15}). Although the broad pattern is similar, there is a small discrepancy between the inner disc kinematics in the observations (top) compared to our simulations (bottom). We attribute this to the secular change in the disc orientation caused by the torque from the binary. This occurs because orbit R2 is not a stable binary-disc configuration. In particular, the torque from the binary on the outer disc vanishes only when the binary is either planar or perpendicular to the disc. It may therefore be fruitful in future investigations to search for orbits which match the observed data but are consistent with one of these arrangements (see discussion below).




\section{Discussion}
\label{sec:discussion}
 If the observed companion can explain the otherwise unexplained features observed in HD142527 disc, this has potential implications for our understanding of mm-bright transition discs in general \citep[c.f.][]{casassus16,owen16}. These represent the fraction of transition discs with large, mm-bright cavities and high mass accretion rates. Observations of transition discs with mass flow onto the central star continuing unabated despite the presence of a large cavity already led numerous authors to conclude that the transition disc population is not a homogeneous class \citep{nsm07,alexanderarmitage07,owenclarke12} and that the likely explanation for the mm-bright subclass is the presence of `objects massive enough to alter the accretion flow' \citep{nam15}. Such an explanation was offered as far back as \cite{skrutskieetal90}.

  Cavities in transition discs are thought to be explained by either photoevaporation or companions --- the latter either by dynamical clearing, similar to gap opening, or by having simply used up the dust to form planets. We have shown that the cavity in HD142527 can be satisfyingly explained by dynamical interaction with the observed binary companion, hereby reclassifying it as a circumbinary disc. The streams of gas across the cavity provide a natural explanation for the high accretion rate. Indeed, the average accretion rate in circumbinary discs with $H/R \sim 0.05$ is expected to be similar (within a factor of a few) to that in a disc around a single object \citep[e.g.][]{farrisetal14,rlp16}. In our simulations we measure a mass accretion rate of $\approx 10^{-7} {\rm M}_\odot / {\rm yr}$ onto the primary, consistent with observational estimates of $2 (\pm 1) \times 10^{-7} {\rm M}_\odot / {\rm yr}$ \citep{mendigutiaetal14}.
 
  Interestingly, HD142527 is not the first transition disc to have been reclassified as circumbinary. Similar reclassifications were made for CoKu Tau/4 \citep{irelandkraus08} (inspiring the first part of our title) and CS Cha \citep{espaillatetal07,guentheretal07}. This is largely due to the difficulty in detecting close-in companions. HD142527 demonstrates that it is easy to hide even relatively massive companions inside discs, given that the companion in this case is massive (stellar!) but was only discovered in 2012 due to its close proximity to the central star \citep{billeretal12}. 
  
  In a recent study, \citet{ragusaetal17} found that circumbinary discs could naturally explain the cavity-edge rings, asymmetries and horseshoes seen in ALMA images of mm-bright transition discs including HD135344B, HD142527, SR21, DoAr 44 and Lk H$\alpha$ 330. Stellar mass companions were only required for the largest asymmetries (i.e., in HD142527). Substellar companions still produce large cavities, but with lumpy rings around the cavity edge, rather than horseshoes. They found that the required companion masses were consistent with the observational upper limits for the above discs, with IRS 48 a possible exception.
  
  There are already tentative detections of central companions in several other cases of discs with large cavities, including HD100546 \citep{quanzetal13,brittainetal14,currieetal14,quanzetal15,currieetal15}, LkCa 15 \citep{sallumetal15} and even more intriguingly in MWC758 \citep{reggianietal17}. There are strong hints in other cases (e.g. HD100453; \citealt{wagneretal15}). It does not require a great leap of imagination to suppose that the observed companions may be responsible for the central cavity, spiral arms, and other features in these cases too.
  
  Whether or not our conclusions can be applied beyond the mm-bright subclass is more speculative and beyond the scope of this paper. Certainly stellar mass companions are ruled out in many cases \citep{pottetal10,kohnetal16,ruiz-rodriguezetal16}, nor would they be expected from the known statistics of the field binary population \citep[e.g.][]{raghavanetal10}. Massive planets, however, could perform a similar role in carving central holes in discs, as shown by \citet{ragusaetal17}.
  
 No modelling is perfect, and despite some success, there are a number of remaining caveats to our modelling of HD142527. The main one from our perspective is the secular change in both the binary and outer disc orientation on long timescales ($\gtrsim$100 orbits). That is, the binary and the disc in our simulations are not in a steady configuration. Recent work by \citet{alyetal15}, \citet{martinlubow17} and \citet{zanazzilai18} suggests that the equilibrium configuration for eccentric binaries inclined by more than $\sim$60$^\circ$ is for the disc to align \emph{perpendicular} to the binary (i.e. in a polar alignment). An interesting follow up would be to try to reproduce the disc features with a binary in such an equilibrium configuration, to see whether it can be ruled in or out (a binary orbit with HD142527B at 90$^\circ$ to the outer disc seems possible; see \citetalias{lacouretal16}). An equilibrium configuration is likely given the typical alignment timescale of order one thousand orbital periods \citep{martinlubow17} --- $10^5$ years in HD142527 --- is much shorter than the disc lifetime ($\sim$ Myrs). Using better orbital constraints to determine whether the binary is in or out of equilibrium with the disc could therefore place strong constraints on formation models.
 
 A second caveat is the mismatch in the orientation of the streams seen in HCO+ emission. Better orbital constraints should also help to solve this problem, since there are currently a wide range of possible orbits. The observed orientation of the streamers itself presents a powerful constraint on the orbit.
 
 If large cavities are produced by eccentric massive companions in the disc plane they should be observed, on average, at large projected separations, possibly even embedded in the disc (as appears to be the case in HD100546). Polar orbits could help to solve this coincidence problem since for a face on disc (such as HD142527) the range in projected separation on the sky is much smaller. For the orbits we chose for HD142527, the companion spends 48\%, 43\% and 31\% of its period at projected separations less than 20 au (0.13") for orbits B1, B2 and B3, respectively. In orbits R1, R2 and R3 this fraction is 19\%, 22\% and 15\%, respectively. So the coincidence problem is not too severe with our chosen orbits.
 
 The most valuable observational follow up would thus be to better constrain the binary orbit and the companion mass, since this directly influences the modelling. Better observations of the inner disc similar to recent observations by \citet{avenhausetal17} would be very valuable in helping to constrain its geometry, mass and orientation. Finally, we found that kinematic data provides a rich source of information on the cavity dynamic. In particular, HCO+ is a powerful probe of the streams of gas crossing the cavity. Detection of similar streamers in other mm-bright transitional discs would be a powerful way to infer the presence of hidden companions.
 
 
 
 That such eccentric and inclined binaries appear to exist in nature opens many fruitful avenues for theoretical investigation into warped and inclined discs around binaries. For example, \citet{owenlai17} proposed that large misalignments of inner and outer discs as in HD142527 may occur through a resonance between the precession period of the inner disc and the precession of the secondary. However, they predict that the orbital plane of the binary should be aligned with the outer disc, which does not appear to be true in HD142527 based on \citetalias{lacouretal16} and our models.
 
 Another interesting direction (in our view) would be to try to infer the binary orbit from the spiral pattern induced around the cavity. This should be possible given sufficient observational constraints on the temperature structure of the disc. Current analytic prescriptions for spiral arms from companions  work only in the linear regime (i.e. for low mass companions; see \citealt{ogilvielubow02,rafikov02}), leading to potentially misleading conclusions regarding the number and mass of the required companions \citep[e.g.][]{stolkeretal16}.
 
 The dynamical and thermodynamical influence of the circumprimary disc shadow on the outer disc in HD142527 also presents an interesting avenue for further investigation. For example, \citet{montesinosetal16} showed that shadows can trigger additional spiral arms in the outer disc. 
  
\section{Conclusions}
\label{sec:summary}

\begin{enumerate}
\item The cavity, spiral arms, shadows, dust horseshoe, gap-crossing filaments and fast radial flows seen in HD142527 can all be explained, in part or in full, by the interaction with the observed central binary companion. HD142527 should therefore be firmly reclassified as a circumbinary rather than transitional disc.
\item Orbits drawn from the best fitting orbits considered by \citetalias{lacouretal16} readily produce a cavity of the required size in HD142527, implying that the observed binary is likely the origin of the large $\approx$ 90~au dust cavity in this disc. Constraints from the cavity size imply a binary semi-major axis $a \lesssim 50$ au.
\item Comparison of the spiral structure and shadows with the observations favours the `red' family of orbits considered by \citetalias{lacouretal16} with the binary approaching periastron, with large eccentricity $e=0.6$--$0.7$, and almost polar inclination with respect to the outer disc. 
\item Binary orbits from \citetalias{lacouretal16} with the companion approaching periastron naturally produce an inclined circumprimary disc with major axis oriented north-south, fed by streams from the outer disc. This orientation of the inner disc is consistent with the radiative transfer model used by \citet{mpc15} to fit the shadows.
\item We find radial velocities across the cavity of order 10 km/s, consistent with the observed `fast radial flows'. We thus offer an explanation for fast radial flows in terms of the streams of material connecting the inner and outer discs.
\item We find gas and dust are decoupled in HD142527. Dust migration to the cavity edge produces features consistent with the observed dust emission, including the prominent mm-horseshoe.
\end{enumerate}

%

 Given that all of the features present in HD142527 are present in some or all mm-bright transition discs, explaining them may help to explain this class of discs in general. For example, spirals and shadows seen in scattered light around the $\sim 45$ au dust cavity in the disc of HD135344B \citep{garufietal13} are best explained by an `inner dust ring' inclined by 22$^{\circ}$ and an `accretion funnel flow' onto the star \citep{stolkeretal16}. In the context of the model we have presented in this paper, these phrases sound eerily familiar.

\section*{Acknowledgements}
  This project was initiated during the workshop on `Planet formation in the era of ALMA and extreme AO' in Santiago, Chile. Section~\ref{sec:observations} documents a lengthy discussion between many of us while sharing a sushi lunch. DJP thanks S. Casassus, J. Cuadra, Universidad de Chile, Pontificia Universidad Cat\'olica de Chile and the Millenium Nucleus for their hospitality and financial support during two visits to Santiago. We thank Yann Boehler, Rebecca Martin and Rebecca Nealon for useful discussions. DJP is funded by an Australian Research Council Future Fellowship FT13010034 and Discovery Projects DP130102078 and DP180104235. We acknowledge CPU time on Gstar/SwinStar at Swinburne University, funded by the Australian Government, and on the MonARCH cluster at Monash. NC acknowledges financial support from FONDECYT grant 3170680. JC and NC acknowledge Millenium Nucleus grant RC130007 (Chilean Ministry of Economy). GMK is supported by the Royal Society as a University Research Fellow. JC acknowledges support from CONICYT-Chile through FONDECYT (1141175) and Basal (PFB0609) grants. We thank the anonymous referee for comments which have improved the manuscript.

\bibliography{dan}

%
%
%
%

\label{lastpage}
\end{document}